\documentclass{ws}
\usepackage{times,epsfig} 

\usepackage{amsmath}
\usepackage{amsthm}
\usepackage{amsfonts}
\usepackage{amssymb}
\usepackage[mathcal]{euscript}
\usepackage{epic}
\usepackage{eepic}
\usepackage{times}
\usepackage{graphics}

\def\1{\one}
\def\2{\two}

\def\H3p{H_3^+}

\newcommand{\SU}{{\rm SU(2)}}

\theoremstyle{plain}
\newtheorem{thm}{Theorem}

\theoremstyle{remark}

\renewcommand{\sst}{\scriptscriptstyle}

\newcommand{\beq}{\begin{equation}}
\newcommand{\eeq}{\end{equation}}

\newcommand{\id}{{\rm id}}

\newcommand{\pa}{\partial}
\newcommand{\ot}{\otimes}
\renewcommand{\ra}{\rightarrow}

\newcommand{\fr}[2]{{\textstyle \frac{#1}{#2} }}

\newcommand{\fsl}{{\mathfrak s}{\mathfrak l}}

\newcommand{\bra}{\langle}
\newcommand{\bbra}{\bra\!\bra}

\newcommand{\ket}{\rangle}
\newcommand{\kket}{\ket\!\ket}

\renewcommand{\al}{\alpha}

\renewcommand{\be}{\beta}
\newcommand{\ga}{\gamma}
\newcommand{\Ga}{\Gamma}
\newcommand{\de}{\delta}
\newcommand{\De}{\Delta}
\newcommand{\ep}{\epsilon}
\newcommand{\la}{\lambda}

\newcommand{\Om}{\Omega}
\newcommand{\si}{\sigma}

\newcommand{\vf}{\varphi}

\newcommand{\ba}{\bar{a}}

\newcommand{\bw}{\bar{w}}

\newcommand{\bz}{\bar{z}}

\newcommand{\CF}{{\mathcal F}}

\newcommand{\CH}{{\mathcal H}}

\newcommand{\CU}{{\mathcal U}}

\newcommand{\SE}{{\mathsf E}}

\newcommand{\SL}{{\mathsf L}}

\newcommand{\SN}{{\mathsf N}}
\newcommand{\SO}{{\mathsf O}}  
\newcommand{\SP}{{\mathsf P}}  
\newcommand{\SQ}{{\mathsf Q}}  

\renewcommand{\SS}{{\mathsf S}}
\newcommand{\ST}{{\mathsf T}}
\renewcommand{\SU}{{\mathsf U}}
\newcommand{\SV}{{\mathsf V}}
\newcommand{\SW}{{\mathsf W}}

\newcommand{\sa}{{\mathsf a}}

\renewcommand{\sf}{{\mathsf f}}
\newcommand{\sh}{{\mathsf h}}

\newcommand{\sq}{{\mathsf q}}
\newcommand{\spp}{{\mathsf p}}
\newcommand{\sr}{{\mathsf r}}
\newcommand{\mss}{{\mathsf s}}

\newcommand{\sx}{{\mathsf x}}

\newcommand{\one}{{\mathfrak 1}}
\newcommand{\two}{{\mathfrak 2}}

\newcommand{\BR}{{\mathbb R}}

\newcommand{\BC}{{\mathbb C}}

\newcommand{\BT}{{\mathbb T}}
\newcommand{\BZ}{{\mathbb Z}}

\renewcommand{\=}[1]{\stackrel{(\ref{#1})}{=}}

\DeclareMathOperator{\sgn}{sgn}

\renewcommand{\ba}{\bar{\mathsf a}}
\newcommand{\rf}[1]{(\ref{#1})}

\newcommand{\aufz}
{\begin{list}{$\bullet$}{\topsep0cm \itemsep0cm \parsep0cm}}
\newcommand{\eaufz}{\end{list}}
\begin{document}
\title{A lecture on the Liouville vertex operators}
\author{J. Teschner}
\address{Institut f\"ur theoretische Physik\\                 
Freie Universit\"at Berlin,\\                        
Arnimallee 14\\                                    
14195 Berlin\\ Germany}
\maketitle
\abstracts{
We reconsider the construction of exponential fields in 
the quantized Liouville theory. It is based on a free-field
construction of a continuous family or chiral vertex operators.
We derive the fusion and braid relations of the chiral vertex 
operators. This allows us to simplify the verification of locality 
and crossing symmetry of the exponential fields considerably. 
The calculation of the matrix elements of the 
exponential fields leads to a constructive derivation
of the formula proposed by Dorn/Otto and the brothers
Zamolodchikov.}
 
\begin{center}
\medskip
\em{Dedicated to A.A.~Belavin on his $60^{\rm th}$ birthday} 
\medskip
\end{center}

\section{Introduction}

Thanks to conformal symmetry it suffices to know 
the three point functions of the exponential fields in 
quantum Liouville theory in order to characterize it completely \cite{BPZ}.  
An explicit formula for these three point functions
was proposed in \cite{DO,ZZ}. This proposal has
successfully passed many checks, giving strong evidence for the
truth of the conjecture. This formula can be considered to encode
the exact solution of quantum Liouville theory once it has
been shown to lead to a solution of the basic consistency requirements
like the crossing symmetry \cite{TL}.

A problem of fundamental importance is to describe  
the monodromies
of the conformal blocks in a conformal field theory. 
Knowing the monodromies not only allows one to
formulate and hopefully solve the basic consistency conditions 
that the three point functions have to satisfy. It also 
forms the basis for a similar approach to the determination
of the structure functions that characterize Liouville theory in 
the presence of a conformally invariant boundary
\cite{PT3}.

A program aimed at the determination of the monodromies
of the conformal blocks in quantum Liouville theory was started in
\cite{PT1}. There we assumed 
that the monodromies can be described 
with the help of a generalization of the Moore-Seiberg formalism for 
rational conformal field theories. Within this formalism it 
suffices to know certain data, called fusion- and braid coefficients, 
in order to reconstruct the monodromies of conformal blocks in
general. These data can be calculated directly in certain special cases.
This information allows one to derive a closed system of functional
equations for the general fusion- and braid coefficients from the
basic consistency conditions of the Moore-Seiberg formalism.
A solution of these functional equations was constructed in 
\cite{PT1,PT2} from the representation theory of the 
modular double of $\CU_{q}(\fsl(2,\BR))$ \cite{Fa,BT}.

A more direct approach was initiated in \cite{TL}. 
Chiral vertex operators were constructed from a free chiral field.
These vertex operators were shown to satisfy exchange relations
which lead to a direct calculation of the fusion- and braiding coefficients.
The result was found to be equivalent to the one from \cite{PT1}.

Unfortunately, the presentation in \cite{TL} was rather brief.
The aim of the present paper will be to explain the constructions
from \cite{TL} in more details and to present the derivations of
some important results that were stated in \cite{TL} without a proof.
In particular, we will derive a formula for the matrix elements 
of the chiral vertex operators. A simplified derivation
of the exchange relations of the chiral vertex operators is presented
here in some detail. With the help of these exchange relations we
will furthermore construct the operator product expansion of the chiral 
vertex operators (fusion).
 
We may then construct the Liouville vertex operators $\SV_{\al}(z,\bz)$ out
of the chiral vertex operators. With the help of the exchange relations
it becomes rather simple to prove the locality 
of the Liouville vertex operators.
Our formula for the matrix elements of the
chiral vertex operators finally allows us to compute the 
matrix elements of the $\SV_{\al}(z,\bz)$. 
Taken together our results represent a constructive
derivation of the formula proposed by Dorn/Otto and the brothers
Zamolodchikov for the three-point function in quantum Liouville 
theory.

The mathematically oriented reader will notice that our treatment
of the vertex operators is somewhat formal. However, we will indicate
in Section 2 that the problems associated with the 
field-theoretical nature of the chiral vertex operators are not worse
than usual. Known techniques will be applicable. What is nonstandard
in our case are the issues associated with the continuous spectrum of the 
zero modes, which lead e.g. to the first example for exchange relations that
involve a continuous set of fields. These are the issues we will mainly be
concerned with.

\section{Definition and main properties of chiral vertex operators}
\label{intI}

\subsection{The chiral free field}

Given the possibility to map classical Liouville theory to a free 
field theory, see e.g. Section 13 of \cite{TL} for a review, 
it is natural to approach quantization
of Liouville theory by first quantizing the free field theory
and then trying to reconstruct the Liouville field operators
in terms of operators in the free field theory.

Let us introduce the (left-moving) 
chiral free field $\vf(x_+)\;=\;\sq+\spp x_+ + 
\vf^{}_<(x_{+})+\vf^{}_>(x_{+}),$ with mode-expansion given by
\begin{equation}
\vf_{<}(x_+)\;=\; 
i\sum_{n< 0}
\frac{1}{n}\, \sa_n \, e^{-inx_+},
\qquad
\vf_{>}(x_+)\;=\;
i\sum_{n> 0}
\frac{1}{n}\,\sa_n \,e^{-inx_+},
\end{equation}
The modes are postulated to have the following 
commutation and hermiticity relations
\begin{equation}\label{ccr}
[\sq,\spp]=\frac{i}{2},\quad
\begin{aligned}\sq^{\dagger}=& \sq,\\
\spp^{\dagger}=&\spp,
\end{aligned}
\qquad [\sa_n,\sa_m]=\frac{n}{2}\de_{n+m},\quad
\sa_n^{\dagger}=\sa_{-n}, 
\end{equation}
which are realized in the  
Hilbert-space
\begin{equation}
\CH^{\rm\sst F}_{\rm\sst L}\;\equiv \;L^2(\BR)\ot\CF,
\end{equation}
where $\CF$
is the Fock-space generated by acting with the modes $\sa_n$, 
$n<0$ on the Fock-vacuum $\Om$
that satisfies $\sa_n\Om=0$,
$n>0$. We will mainly work in a representation where $\spp$ is diagonal.

\subsection{Conformal symmetry}\label{confsym}

A representation of the 
conformal symmetry is defined on $\CH^{\rm\sst F}_{\rm\sst L}$ by means of
the standard free field realization of the Virasoro algebra.
The action of the Virasoro algebra on $\CH^{\rm\sst F}_{\rm\sst L}$ can 
be defined in terms
of the generators $\SL_n\equiv\SL_n(\spp)$, where
\begin{equation}\label{FockVir}
\begin{aligned}\SL_n^{}(p)\;=\;& 
(2p+inQ)\sa_n^{}+\sum_{k\neq 0,n}\sa_k^{} \sa_{n-k}^{},
\qquad n\neq 0, \\
\SL_0^{}(p)\;=\;&p^2 +\frac{Q^2}{4}+
2\sum_{k>0}\sa_{-k}^{}\sa_k^{}. 
\end{aligned}
\end{equation}
Equations \rf{FockVir} are well-known to yield a representation
of the Virasoro algebra with central charge
\begin{equation}
c\;=\; 1+6Q^2.
\end{equation}
We will mostly consider the case that $Q>2$ in the following, corresponding 
to central charge $c>25$. It turns out, however, that the results 
that we obtain for this regime have
an analytic continuation w.r.t. the parameter $Q$ 
which allows one to cover the case $c>1$ as well.  

Equations \rf{FockVir} define a one-parameter family of representations
of the Virasoro algebra.
This means that $\CH^{\rm\sst F}_{\rm\sst L}$ 
decomposes as the direct integral of
Fock-space representations of the Virasoro algebra:
\begin{equation}\label{decFP}
\CH^{\rm\sst F}_{\rm\sst L}\;\simeq\;\, \int\limits_{-\infty}^{\infty}dp
\;\CF_p\;,
\end{equation}
where $\CF_p$  denotes the Virasoro representations
defined on $\CF$ by means of the generators $\SL_n^{}(p)$ 
defined in \rf{FockVir}. The representations 
$\CF_p$, are known to be unitary
highest weight representations of the Virasoro algebra, the weight
being given in terms of $p$ via $\De_p=p^2+\frac{1}{4}Q^2$.
We will denote the highest weight vector in $\CF_p$ by $v_p$.

The representations $\CF_p$ and $\CF_{-p}$
are unitarily equivalent. This follows immediately 
from the fact \cite{Fr} that the monomials 
\[
\SL_{-m_\1}(p)\dots \SL_{-m_l}(p)\Omega
\] 
form a basis for $\CF$ for any $p\in\BR$. Replacing $p\ra -p$
therefore defines a unitary map
$\SS(p):\CF_p\ra
\CF_{-p}$. It is useful to collect the 
maps $\SS(p)$, $p\in \BR$ into a single operator 
$\SS\equiv\SS(\spp):\CH^{\rm\sst F}_{\rm\sst L}\ra 
\CH^{\rm\sst F}_{\rm\sst L}$.

The representations are known to exponentiate to 
projective unitary representations of the group ${\rm Diff}(S_\1)$
of orientation-preserving
diffeomorphisms of the circle. Elements of ${\rm Diff}(S_\1)$ can be
parameterized by monotonic smooth 
functions $h:\BR\ra\BR$ that satisfy $h(\si+2\pi)=h(\si)+2\pi$. We will
use the notation $\SU_h$ 
for the operator that represent 
the diffeomorphism $h$ in the representations of ${\rm Diff}(S_\1)$
generated by the $\SL_n$.

\subsection{Building blocks}

The basic building blocks of all constructions will be the following 
objects:\\[1ex]
\noindent{\sc Normal ordered exponentials : }
\begin{equation} 
\SE^{\al}(x_{+})\;
\equiv\;
\SE^{\al}_{<}(x_{+})\SE^{\al}_{>}(x_{+}),
\qquad 
\begin{aligned}
{} & 
\SE^{\al}_<(x_{+})\;=\;
e^{\al \sq}\;
e^{2\al\vf^{{+}}_{<}(x_{+})}\;
e^{\al x_+ \spp}\\
{} & 
\SE^{\al}_>(x_{+})\;=\;
e^{\al x_+ \spp}\;
e^{2\al\vf^{{+}}_{>}(x_{+})}\;
e^{\al\sq} 
\end{aligned}
\end{equation}
{\sc Screening charges:}\footnote{In comparison with \cite{TL} we have
modified our definition by the factors $e^{-\pi b\spp}$. This allows
us to get rid of some annoying exponential prefactors in \cite{TL}.}
\begin{equation}
\SQ(x)\;\equiv\;e^{-\pi b\spp}
\int\limits_0^{2\pi}d\si\;\SE^b(x+\si)\;e^{-\pi b\spp}\;.
\end{equation}
It will be quite important for our purposes to understand the 
mathematical nature of these objects a little better.
Let us first of all recall that fields like 
the normal ordered exponentials do {\it not} represent true operators. 
Indeed, due to the usual short-distance
singularities we can never make sense out of 
$\lVert \SE^{\al}(x_+)\psi\rVert^2$. There are two standard ways to
treat this nuisance: On the one hand one may systematically use {\it smeared} 
fields, obtained by multiplying with some test-function and integrating 
over the cylinder. Alternatively one may analytically continue to 
negative Euclidean time, i.e.
assume that $\Im(x_+)>0$. 
Objects like $\SE^{\al}(\si-i\tau)$, 
$\tau<0$ will indeed represent
densely defined, but unbounded operators. Combinations like
\[
\Pi_p\SE^{\al}(\si-i\tau)e^{-\ep\SL_0} :\CF_{p+i\al}\ra\CF_p,
\] 
where $\ep>0>\tau$ and $\Pi_p$ denotes
the projection onto $\CF_p$, are even trace-class,
as can be shown by means of a direct calculation of the trace in a
coherent-state basis for $\CF$. 

The screening charges on the contrary make sense as
true operators even for $x\in\BR$ 
as their definition already involves some smearing. 
More precisely: Let $2b^2<1$. The screening charges $\SQ(\si)$
then represent densely defined unbounded operators. 
To verify this statement let us note 
that 
\[
\big\lVert Q(\si)\psi \big\rVert^2 \; =\;
\int\limits_{\si}^{2\pi+\si}\!\! d\si' d\si''\;
\bigl\langle \psi, Q(\si')Q(\si'')\psi\bigr\rangle.
\]
The integrand develops for $\si'\ra \si''$ a singularity of the form 
$|\si'-\si''|^{-2b^2}$ which is integrable for $2b^2<1$. 
Other values of $b$ can be covered by means of analytic continuation
w.r.t. $b$. 

Let us note the following
crucial property:\\[1ex]
{\sc Positivity: } The screening charges are positive operators.
Indeed, we have that
\[
\langle \psi, Q(\si)\psi\rangle\;=\;
\int\limits_{\si}^{2\pi+\si}\!\! d\si' \; 
\big\lVert \,\SE^b_>(\si')\psi\,
\big\rVert^2 \;>\;0\;.
\]
This property will be of fundamental importance for the
rest since our construction of Liouville vertex operators
will involve {\it complex} powers $s$ of the screening charges.
Positivity of $\SQ(\si)$  allows one to take 
arbitrary powers of these operators.

\subsection{Chiral vertex operators}

Out of the building blocks introduced in the previous subsection 
we may now construct an important class of chiral fields.
\begin{equation}
\sh_s^{\al}(x_+)\;=\;
\SE^{\al}(x_+)\,
\big(\SQ(x_+)\big)^s
\;,\end{equation}
Positivity of $\SQ$ allows us to consider these
objects for {\it complex} values of $s$ and $\al$. 

The chiral fields $\sh_s^{\al}(x_+)$ will 
have similar analytic properties as the normal ordered
exponentials. As usual it is most convenient to work
with the Euclidean fields obtained by analytically continuing to
imaginary time. To be specific, let us define 
the Euclidean fields $\sh_s^{\al}(w)$, $w=\tau+i\si$ on 
the domain of $e^{-\tau \SL_0}$ as
\begin{equation}\label{euclfield}
\begin{aligned}
{}& \sh_s^{\al}(w)\;\equiv\;e^{(\tau+\ep)\SL_0}\,\sh_{s,\ep}^{\al}(\si)\,
e^{-\tau \SL_0},\\
& \sh_{s,\ep}^{\al}(\si)\;\equiv\; \SE^{\al}(\si+i\ep)\,
 e^{-\ep\SL_0}\big(\SQ(\si)\big)^s,
\end{aligned}\qquad \tau<0<\ep<|\tau|.
\end{equation}  

One of the most basic properties of the $\sh_s^{\al}(w)$ are the
simple commutation relations with functions of $\spp$,
\begin{equation}\label{shift}
\sh_s^{\al}(w)f(\spp)\;=\;f\big(\spp-i(\al+bs)\big)
\sh_s^{\al}(w),
\end{equation}
which follow from the fact that $\sh_s^{\al}(w)$ depends on $\sq$
only via an overall factor $e^{2(\al+bs)\sq}$.

It is often convenient to work with $\Pi_p\sh_s^{\al}(w)$
rather than $\sh_s^{\al}(w)$, where $\Pi_p$ denotes the
projection onto an eigenspace $\CF_p\simeq \CF$ of
$\spp$. Equation \rf{shift} implies that 
$\Pi_p\sh_s^{\al}(w)$ projects onto $\CF_q$, where $q=p+i(\al+bs)$.
In order to assure $q\in\BR$ we will mostly assume $i(\al+bs)\in\BR$ in 
the following. More general cases can afterwards be treated by analytic
continuation.
$\Pi_p\sh_s^{\al}(w)$ therefore defines a family of operators
\begin{equation}
\sh_{p_\2p_\1}^{\al}(w): \CF_{p_\1}\ra\CF_{p_\2}.
\end{equation} 
The operators $\sh_{p_\2p_\1}^{\al}(w)$ are called chiral
vertex operators. 

We would like to study products of these operators such as
$\sh_n(w_n)\dots\sh_\1(w_\1)$, $\sh_k(w_k)\equiv
\sh_{p_kp_{k-1}}^{\al_k}(w_k)$ and their matrix elements. 
In order to convince ourselves that these objects are well-defined
let us observe that by projecting $\sh_{s,\ep}^{\al}(\si)$, cf. 
(\ref{euclfield}), onto $\CF_p$ one gets an operator 
$\sh_{p_\2p_\1}^{\al,\ep}(\si)$ that is the product of a trace-class 
operator with an operator that is bounded for $s\in i\BR$. With the 
help of (\ref{euclfield}) one 
may then represent the matrix elements of 
$\sh_n(w_n)\dots\sh_\1(w_\1)$, 
in the following form: 
\[ \begin{aligned}
e^{(\tau_n+\ep)\De_{p_n}-\tau_\1\De_{p_0}}
\big\langle  v_{p_n}\, ,\, 
\sh_{n,\ep}(\si_n)\, & e^{-(\tau_n-\tau_{n-1}-\ep)\SL_0}\dots\\
& \dots \sh_{2,\ep}(\si_\2)
e^{-(\tau_\2-\tau_\1-\ep)\SL_0}\,\sh_{1,\ep}(\si_\1)\,
v_{p_0} \big\rangle \; ,
\end{aligned} \]
where we have set $\sh_{k,\ep}(\si_k)\equiv
\sh_{p_kp_{k-1}}^{\al_k,\ep}(\si_k)$, $k=1,\dots,n$.
These matrix elements are analytic w.r.t. the variables 
$\tau_k-\tau_{k-1}\in{\mathbb H}_+$, 
where ${\mathbb H}_+=\{z\in\BC;\Re(z)>0\}$.
Indeed, the positivity of $\SL_0$
implies analyticity of the vector-valued function 
$e^{-\tau \SL_0}\xi$, $\xi\in\CH^{\rm\sst F}_{\rm\sst L}$ 
on the right $\tau$-half-plane.

The convergence of the power series expansions in 
$q_k=e^{\tau_{k}-\tau_{k-1}}$
finally allows one to establish the meromorphic continuation of the 
matrix elements to more general values of $\al_k$, $k=1,\dots,n$ 
and $p_l$, $l=0,1,\dots,n$, cf. \cite{TL}, Section 7.

\subsection{Conformal covariance}

The behavior of the normal ordered exponentials 
$\SE^{\al}$
under conformal transformations can then be 
summarized by 
\begin{equation}\label{Evir}
\SU_h\,\SE^{\al}(\si)\,\SU^{\dagger}_h\;=\;\bigl(
h'(\si) 
\bigr)^{\De_{\al}}\,\SE^{\al}\big(h(\si)\big),\qquad
\De_{\al}=\al(Q-\al).
\end{equation}
Let us now assume that the parameter $b$ that enters the 
definition of the
screening charge $\SQ$
is related to the parameter $Q$ via 
\begin{equation}
Q=b+b^{-1}.
\end{equation}
In this case \rf{Evir} implies the following 
simple transformation law for $\SQ(\si)$: 
\begin{equation}\label{Qvir}
\SU_h\,\SQ(\si)\,\SU^{\dagger}_h\;=\;\SQ\big(h(\si)\big).
\end{equation}
Let us note that this implies the true invariance of $\SQ(\si)$ under those
elements of ${\rm Diff}(S_\1)$ that satisfy $h(\si)=\si$.
Equations \rf{Evir} and \rf {Qvir} together finally imply that
\begin{equation}\label{gvir}
\SU_h\,\sh^{\al}_s(\si)\,\SU^{\dagger}_h\;=\;\bigl(
h'(\si) 
\bigr)^{\De_{\al}}\,\sh^{\al}_s\big(h(\si)\big),\qquad
\De_{\al}=\al(Q-\al),
\end{equation}
which is the standard transformation law for a covariant chiral 
field.

It is often convenient to trade the Euclidean cylinder
parameterized by the coordinate $w$ for the complex plane by
making the conformal transformation $z=e^w$. The corresponding 
fields on the complex plane will simply be denoted by
$\sh_s^{\al}(z)$. These fields are related to the $\sh_s^{\al}(w)$
via
\begin{equation}\label{cylplane}
\sh_s^{\al}(w)\;=\;z^{\De_{\al}}\,\sh_s^{\al}(z).
\end{equation} 
The composition of two 
fields $\sh_{s_\2}^{\al_\2}(z_\2)\sh_{s_\1}^{\al_\1}(z_\1)$ 
will as usual be well-defined
if $|z_\2|>|z_\1|$. 

Having discussed their mutual relations 
we shall in the following freely switch between
distributional covariant fields $\sh^{\al}(\si)$ on the 
(universal cover of the) unit circle 
$S_\1$, the corresponding 
operators $\sh^{\al}(w)$ on the Euclidean cylinder  and 
their counterparts $\sh^{\al}(z)$ on the punctured complex plane
which are obtained via \rf{cylplane}.

\subsection{More advanced properties of $\sh_{p_\2p_\1}^{\al}(w)$}

Let us now summarize 
the properties of the fields $\sh_{p_\2p_\1}^{\al}(w)$ that will
be crucial for their role as building blocks in the construction 
of Liouville vertex operators. Derivations will be given 
in the following sections.\\[1ex]
{\sc Matrix elements} $\frac{\quad}{}$
\begin{equation}\label{g-matel}
\begin{aligned}
{} & e^{w(\De_{p_\1}-\De_{p_\2})}\;
\langle  v_{p_\2}\, ,\, \sh_{p_\2p_\1}^{\al}(w)\,
v_{p_\1} \rangle\;=\;\\
& \qquad =\;  \big(\Ga(b^2)b^{1-b^2}\big)^s\;
\frac{|\, \Ga_b(Q-\al+i(p_\2+p_\1))
\Ga_b(Q-\al+i(p_\2-p_\1))\,|^2}{\Ga_b(Q)
\Ga_b(Q-2ip_\2)\Ga_b(Q-2\al)\Ga_b(Q+2ip_\1)}
\end{aligned}
\end{equation}
We have assumed $\al\in\BR$ in order to write the 
expression compactly. The 
definition and some relevant properties of the 
special function $\Ga_b(x)$ \cite{Ba} are collected in the Appendix.\\[1ex]
{\sc Exchange relations} $\frac{\quad}{}$
\begin{equation}\label{braid}
\sh_{p_\2p_s}^{\al_\2}(\si_{\2})\,
\sh_{p_sp_\1}^{\al_\1}(\si_{\1})\;=\;
\int\limits_{0}^{\infty}dp_u \;\,B_{\rm\sst E}^{\ep}(\,p_s\,|\,p_u)\;
\sh_{p_\2p_u}^{\al_\1}(\si_{\1})\,\sh_{p_up_\1}^{\al_\2}(\si_{\2}),
\end{equation}
where $\ep\equiv \sgn(\si_\2-\si_\1)$.
In the notation for the kernel
$B_{\rm\sst E}^{\ep}$ we have denoted the tuple of external parameters by
${\rm E}=(\al_\2,\al_\1,p_\2,p_\1)$. The explicit expression for 
$B_{\rm\sst E}^{\ep}(\,p_s\,|\,p_u)$ is of the following form:
\begin{equation}\label{braidcoeff1}\begin{aligned}
B_{\rm\sst E}^{\ep}(\,p_s\,|\,p_u)\;=\;&
e^{\pi i\ep(\De_s+\De_u-\De_{\2}-\De_{\1})}
B_{\rm\sst E}(\,p_s\,|\,p_u), \\
B_{\rm\sst E}(\,p_s\,|\,p_u)\;=\;&
{\rm m}(p_u)
\int\limits_{\BR+i0}dt\;\,\prod_{i=1}^4\;\frac{s_b(t+r_i)}{s_b(t+s_i)},
\end{aligned}\end{equation}
where ${\rm m}(p)=
4\sinh2\pi bp\sinh2\pi b^{-1}p$,
$\De_{\natural}=p_{\natural}^2+\frac{Q^2}{4}$, $\flat\in\{1,2,s,u\}$,  
and the coefficients $r_i$ and $s_i$, $i=1,\ldots,4$ are defined by
\begin{equation}  
\begin{aligned} r_\1=& i(\al_\2-\al_\1)-p_\2,\\
         r_\2=& i(\al_\2-\al_\1)+p_\2, \\
         r_3=& +p_\1,\\
        r_4=& -p_\1,
\end{aligned}\qquad\qquad
\begin{aligned} 
       s_\1=& +p_u-i(Q-\al_\2),\\
         s_\2=& -p_u-i(Q-\al_\2),\\
         s_3=& +p_s-i\al_\1,\\
        s_4=& -p_s-i\al_\1. 
\end{aligned}
\end{equation}
The special function $s_b(x)$ can be defined by 
$s_b(x)=\Ga_b\big(\frac{Q}{2}+ix\big)\,/\,
\Ga_b\big(\frac{Q}{2}-ix\big)$, see the
Appendix for further information.
For our purposes the most important property of the kernel 
$B_{\rm\sst E}^{\ep}(\,p_s\,|\,p_u)$
will be  the unitarity relation 
\begin{equation}\label{unit}
\int\limits_{0}^{\infty}dp_s\,{\rm m}(p_s)\;
B_{\rm\sst E}^{\ep}(\,p_s\,|\,p_u)\,
\big(B_{\rm\sst E}^{\ep}(\,p_s\,|\,p_u')\big)^{*}\;=\;
{\rm m}(p_u)\,\de(p_u-p_u')\; .
\end{equation}
The relations \rf{braid} and \rf{unit} will be the main ingredients 
that one needs to 
prove the locality of Liouville vertex operators.

\section{Matrix elements}

We shall present here the derivation of formula \rf{g-matel} 
for the matrix elements of the fields 
$\sh_s^{\al}$. Let us define the function
$M(\al,s\,|\,p)$ as 
\begin{equation}
M(\al,s\,|\,p)\;\equiv\;
\langle\!\langle \, p'\,|\,\sh_s^{\al}(1)\,
|\,p\, \rangle\!\rangle ,\qquad p'\equiv p-i(\al+bs),
\end{equation}
where we have used the notation
\begin{equation}
\langle\!\langle \, p_\2\,|\,\SO(w)
\,|\,p_\1 \,\rangle\!\rangle\;=\;\bigl( \,v_{p_\2}\, , \,
\Pi_{{p_\2}}\SO(w)\,v_{p_\1}\big)_{\CF}\;.
\end{equation}
Let us emphasize that
we are talking about a perfectly well-defined object, our task is just
to make it more explicit. What we are going to assume, but not prove 
here, will be that $M(\al,s\,|\,p)$ has a sufficiently large domain
of analyticity in its dependence w.r.t. $\al$ and $s$. 
Our strategy will be to derive a system of  
functional equations that 
will then
characterize the function $M(\al,s\,|\,p)$ completely. 


\subsection{Auxiliary fields}

As a useful technical device we shall employ the
following set of fields:
\begin{equation}\label{auxfields}
\begin{aligned}
\sh_0(z)\; \equiv & \;\sh^{-\frac{b}{2}}_0(z)\;=\;\SE^{-\frac{b}{2}}(z),
\\
\sh_\1(z)\; \equiv \;& \;\sh^{-\frac{b}{2}}_\1(z)\;=\;
e^{\frac{\pi i}{2}b^2}
e^{-\pi b\spp}\SE^{-\frac{b}{2}}(z)\SQ(z)\,e^{-\pi b\spp}\;. 
\end{aligned}
\end{equation}
We will need the following two properties of the fields $\sh_r(z)$, 
$r=0,1$:\\[1ex]
{\sc Matrix elements:}
\begin{equation}\label{m_el}\begin{aligned}
M_0\;\equiv\;&\bbra \,p+i\fr{b}{2}\,|\,\sh_0(1)\,|\,p\,\kket\;=\;1\\
M_\1\;\equiv\;&\bbra \,p-i\fr{b}{2}\,|\,\sh_\1(1)\,|\,p\,\kket\;=\;2\pi 
\frac{\Ga(1+b^2)}{\Ga(1+b^2+2ibp)\Ga(1-2ibp)}.
\end{aligned}\end{equation}
{\sc Differential equation:} \cite{GN}\cite{BPZ}
\begin{equation}\label{qlin}
 \pa^2  \;\sh_r \;= \;-b^2:\ST\;\sh_r:\;\, ,\quad r=0,1,
\end{equation}
where the normal ordering of the expression on the right hand side 
is defined as follows:
\begin{equation}\label{normord}
{} :\ST\;\sh_r:\;\, = \;\sum_{n\leq -2} \;z^{-n-2}\SL_n \;
\sh_r+\sum_{n\geq -1}
\;\sh_r\;\SL_nz^{-n-2}.
\end{equation}

In the rest of this subsection let us verify equation \rf{m_el}.
The proof of \rf{m_el} is trivial for $\sh_0$.
In the case of $\sh_\1$ we shall use the standard normal ordering formula
\begin{equation}\label{nof} \begin{aligned}
\SE_>^{\al_\2}& (\si_\2)\SE^{\al_\1}_<(\si_\1)
\;=\;\\
&=\;e^{i\al_\2\al_\1(\si_\1-\si_\2)}
\big(1-e^{i(\si_\1-\si_\2)}\big)^{-2\al_\2\al_\1}\,\SE^{\al_\1}_{<}(\si_\1)
\SE^{\al_\2}_{>}(\si_\2)\\
& =\;e^{-\pi i \al_\2\al_\1\sgn(\si_\2-\si_\1)}
|1-e^{i(\si_\1-\si_\2)}|^{-2\al_\2\al_\1}\;
\SE^{\al_\1}_{<}(\si_\1)
\SE^{\al_\2}_{>}(\si_\2).
\end{aligned} \end{equation}
The complex power function in the second line 
of \rf{nof} is defined as $z^s=e^{s\ln(z)}$, where $\ln(z)$ denotes the 
principal value of the logarithm. 
Using \rf{nof}  reduces the calculation of 
$\bbra p-i\fr{b}{2}|\sh_\1(0)|p\kket$ to the integral
\begin{equation}\label{int}
\bbra p-i\fr{b}{2}|\,\sh_\1(0)\,|p\kket\;=\;e^{\pi i b^2}
e^{-2\pi bp}
\int\limits_0^{2\pi}d\si'\; e^{2b\si'(p-i\fr{b}{2})}
\big(1-e^{i\si'}\big)^{b^2}.
\end{equation}
By means of a simple contour 
deformation one easily reduces the evaluation of 
\rf{int} to the integral that 
defines the Beta-function. 

\subsection{Strategy}

Our strategy will be to consider the conformal blocks
\begin{equation}\label{psidef}
\Psi_r(z_\2,z_\1)\;=\;\bbra p_\2|\,\sh_r(z_\2)\,\sh_s^{\al}(z_\1)\,
|p_\1\kket,
\end{equation}
where $p_\2=p_\1-i(\al+bs+\de_r)$ with $\de_0=-\frac{b}{2}$, 
$\de_\1=\frac{b}{2}$.
We will see that the leading singular behavior of
$\Psi_r(z_\2,z_\1)$ for $z_\2\ra z_\1$ is of the form
\begin{equation}
\Psi_r(z_\2,z_\1)\;\underset{z_\2\ra z_\1}{\simeq}
\;(z_\2-z_\1)^{b\al}\,G_r(\al,s,p_\1).
\end{equation}
The coefficient $G_r(\al,s,p)$ can be expressed in terms of 
$M(\al,s\,|\,p)$ in two different ways: 

On the one hand one may calculate
the leading operator product singularity of $\sh_r(z_\2)\,\sh_s^{\al}(z_\1)$
directly from the definition of these operators, with the
result 
\begin{equation}\label{ope}
\sh_r(z_\2)\sh_s^{\al}(z_\1)\;\underset{z_\2\ra z_\1}{\simeq}\;
(z_\2-z_\1)^{b\al}\,\sh_{s+r}^{\al-\frac{b}{2}}(z_\1)\;.
\end{equation}
Indeed, to verify \rf{ope} it suffices to observe 
that 
\[\begin{aligned}
\SQ(\si_\2)\sh^{\al}_s(\si_\1)\;=\; &\SQ(\si_\1)\sh^{\al}_s(\si_\1)+
\big(\SQ(\si_\2)-\SQ(\si_\1)\big)\sh^{\al}(\si_\1)\\
 \underset{\si_\2\ra\si_\1}{\sim} & \sh^{\al}_{s+1}(\si_\1)
\end{aligned}
\]
up to terms that are sub-leading for $\si_\2\ra\si_\1$.
Inserting \rf{ope} into the definition \rf{psidef} of $\Psi_r(z_\2,z_\1)$
yields 
\begin{equation}\label{F-eins}
G_r(\al,s,p_\1)\;=\;M\big(\al-\fr{b}{2},s+r\,|\,p_\1)
\;\, .
\end{equation}

On the other hand one may use the differential equation 
\rf{qlin} to calculate $\Psi_r(z_\2,z_\1)$ explicitly in terms of 
hypergeometric functions. Standard formulae for the asymptotic behavior of
solutions to the hypergeometric differential equation 
will then give a second expression for 
$G_r(\al,s,p)$. Comparison of the two expressions yields the desired 
functional equations.

\subsection{Calculation of the conformal blocks}

We will find it convenient to use the notation
\begin{equation}
\begin{aligned}
\al_\2\equiv& \;-\fr{b}{2},\\
\al_\1\equiv& \;\al,
\end{aligned}\qquad\qquad
\begin{aligned}
\be_\1\equiv& \;\fr{Q}{2}+ip_\1,\\
\be_\2\equiv& \;\fr{Q}{2}+ip_\2 .
\end{aligned}
\end{equation}
Conformal invariance
restricts $\Psi_r(z_\2,z_\1)$ to have the form
\begin{equation}\label{scinv}
\Psi_r(z_\2,z_\1)\;=\;z_\1^{\kappa}\;\Psi_r(z), \qquad
z\equiv \frac{z_\2}{z_\1},\qquad\kappa\equiv\De_{\be_\2}-\De_{\be_\1}
-\De_{\al_\1}-\De_{\al_\2} \;\;.
\end{equation}
The differential equation \rf{qlin} then implies that 
$\Psi_r(z)$ must satisfy the differential equation
\begin{equation}
\bigg(\frac{1}{b^2}\frac{\pa^2}{\pa z^2}+
\frac{2z-1}{z(1-z)}\frac{\pa}{\pa z}+\frac{\De_{\al_\1}}{(1-z)^2}+
\frac{\De_{\be_\1}}{z^2} -\frac{\kappa}{z(1-z)}\bigg)\Psi_r(z)=0\;.
\end{equation}
By making the ansatz 
\begin{equation}\label{hypans}
\Psi_r(z)\;=\;z^{b\be_\1}(1-z)^{b\al_\1}F_r(z)
\end{equation}
one finds the hypergeometric differential equation for $F_r(z)$.
In order to determine the relevant solutions let us observe that 
\begin{equation}
\label{asym1}
\sh^{\al}_s(z_\1)v_p \;\underset{z_\1\ra 0}{\simeq}
z_\1^{\De_{q}-\De_{\al}-\De_{p}}\;
M(\al,s\,|\,p)\,v_{q}, \qquad q=p-i(\al+bs)\;\;.
\end{equation}
Indeed, the limit $z\ra 0$ corresponds to taking $\tau\ra -\infty$
in \rf{euclfield} which obviously suppresses the contributions
of states with higher $\SL_0$-eigenvalues. 
In the present case we need to take $z\ra\infty$ in order to
apply \rf{asym1}. It follows that the asymptotic behavior
of $\Psi_r(z)$ for $z\ra\infty$ must be of the following form:
\begin{equation}\label{asym2} \begin{aligned}
{}& \Psi_r(z)\;\underset{z\ra \infty}{\simeq}\;
\;M_r\;
z^{b(\be_\2-\al_\2)}
z^{rb(Q-2\be_\2)}\;\;,\\
& N_r\;=\; M_r \;M(\al_\1,s\,|\,p_1)\;\;.
\end{aligned}\end{equation}
There exist unique solutions of the hypergeometric differential 
equation that have the required asymptotic behavior
\rf{asym2}, namely
\begin{equation}\label{hypsol}
F_r(z)\;=\; N_r\;z^{-u_r}F\big(u_r,v_r;w_r;\fr{1}{z}\big), 
\end{equation}
where
\begin{equation}
\begin{aligned}
u_0=&\;b(\al_\1+\al_\2+\be_\1-\be_\2),\\
v_0=&\;b(\al_\1+\al_\2+Q-\be_\1-\be_\2),\\
w_0=&\;1-b(2\be_\2-Q),
\end{aligned}
\qquad
\begin{aligned}
u_\1=&\;b(\al_\1+\al_\2+\be_\1+\be_\2-Q),\\
v_\1=&\;b(\al_\1+\al_\2+\be_\2-\be_\1),\\
w_\1=&\;1+b(2\be_\2-Q).
\end{aligned}
\end{equation}
By combining equations \rf{scinv},\rf{hypans} and \rf{hypsol} 
one gets an explicit expression for 
$\Psi_r(z_\2,z_\1)$ in terms of 
the matrix elements $M_r$ and $M(\al,s\,|\,p)$.

\subsection{Functional equations}

The asymptotic behavior of $\Psi_r(z_\2,z_\1)$ can now be calculated with the 
help of the well-known formula
\[ \begin{aligned}
F(u,v;w;z)\;=\;& \frac{\Ga(w)\Ga(w-u-v)}{\Ga(w-u)\Ga(w-v)}
F(u,v;1+u+v-w;1-z)\\
+(1-z)^{w-u-v}& 
\frac{\Ga(w)\Ga(u+v-w)}{\Ga(u)\Ga(v)}
F(w-u,w-v;1+w-u-v;1-z).
\end{aligned}
\]
The result is of the form \rf{ope}, with 
\begin{equation}\label{F-zwei}
G_r(\al,s,p)\;=\;M_r \;M(\al,s\,|\,p)
\frac{\Ga(w_r)\Ga(w_r-u_r-v_r)}{\Ga(w_r-u_r)\Ga(w_r-v_r)}.
\end{equation}
By comparing \rf{F-eins} and \rf{F-zwei} 
and plugging in \rf{m_el} one deduces the 
two functional equations
\begin{equation}\label{funeqs}
M\big(\al-\fr{b}{2},\rho+rb\,|\,\be\big)\;=\;
\chi^{r}_{b}(\al,\rho\,|\,\be)\,
M(\al,\rho\,|\,\be)
\end{equation}
where we have traded the parameter $s$ for $\rho=bs$ and
the parameter $p$ for $\be=\frac{Q}{2}+ip$.
The coefficients $\chi^{r}_{b}(\al,\rho\,|\,p)$, $r=0,1$ are given
by the following expressions:
\begin{equation}
\begin{aligned}
\chi^{0}_{b}(\al,\rho\,|\,\be)\;=\;&
\frac{\Ga(2-b(2\al+2\be+2\rho-2b))\Ga(1-b(2\al-b))}
{\Ga(2-b(2\al+2\be+\rho-2b))\Ga(1-b(2\al+\rho-b))},\\
\chi^{1}_{b}(\al,\rho\,|\,\be)\;=\;&
\frac{2\pi \Ga(1+b^2)\Ga(1-b(2\al-b))}{
\Ga(b(\rho+Q))\Ga(b(2\be+\rho))\Ga(1-b(2\be+2\al+2\rho-Q))}.
\end{aligned}
\end{equation}

Let us furthermore observe that the whole argument leading to \rf{funeqs} 
can be repeated with the alternative set $\tilde{\sh}_r(z)$, $r=0,1$ 
of auxiliary fields that is obtained by replacing $b\ra b^{-1}$
in the definition \rf{auxfields}. This becomes possible due to the fact 
that the fields $\tilde{\sh}_r(z)$, $r=0,1$ satisfy 
\begin{equation}
 \pa^2  \;\tilde{\sh}_r \;= \;-b^{-2}:\ST\;\tilde{\sh}_i:\;\, ,\quad r=0,1.
\end{equation}
As the result one finds a second set of functional equations:
\begin{equation}\label{funeqs'}
M\big(\al-\fr{1}{2b},\rho+\fr{r}{b}\,|\,p\big)\;=
\;\chi^{(r)}_{1/b}(\al,\rho\,|\,p)\,
M(\al,\rho\,|\,p).
\end{equation}

The combined system of functional equations \rf{funeqs}\rf{funeqs'} 
severely constrains the dependence of $M(\al,\rho\,|\,p)$ w.r.t. the variables
$\al$ and $\rho$. 
However, it does not constrain the 
dependence w.r.t. $p$. This freedom is eliminated
by noting that one must have 
\begin{equation}\label{M-norm}
M(\al,\rho\,|\,p)\big|_{\rho=0}\;\equiv\; 1
\end{equation}
for all $\al$, $p$.
We therefore have a system of equations that can be expected to 
characterize $M(\al,\rho\,|\,p)$ completely. 
It certainly does if $b$ is irrational and if 
$M(\al,\rho\,|\,p)$ is known to have a sufficiently large 
domain of analyticity w.r.t. $\al$ and $\rho$, see e.g. \cite{PT3},
Appendix C.

To conclude let us observe that by using 
\rf{Ga_feq}\rf{self-dual} it is straightforward 
to check that our expression
\rf{g-matel} indeed solves the functional equations \rf{funeqs}\rf{funeqs'}.
The condition \rf{M-norm} is trivially fulfilled by \rf{g-matel}.

\section{Braiding}

\subsection{Operator ordering, I}

Let us now study the bilocal field
$\sh_{s_\2}^{\al_\2}(\si_\2)\sh_{s_\1}^{\al_\1}(\si_\1)$ .
In the following we shall concentrate on the 
case $\si_\2>\si_\1$, the case $\si_\1<\si_\2$ being completely analogous.
In order to find a useful alternative representation for this field
we shall reorder the ingredients that enter the definition
of $\sh_{s_\2}^{\al_\2}(\si_\2)\sh_{s_\1}^{\al_\1}(\si_\1)$ conveniently.
To this aim let us introduce the intervals 
$I=[\si_\1,\si_\2]$, $I^c=[\si_\2,\si_\1+2\pi]$ and 
$I'=[\si_\1+2\pi,\si_\2+2\pi]$ together with the corresponding operators
\begin{equation}
\SQ_{\rm\sst I}^{}\;\equiv\;\int_I d\si\;\SE^b(\si),\quad
\SQ_{\rm\sst I}^c{}\;\equiv\;\int_{I^c}d\si\;\SE^b(\si),\quad
\SQ_{\rm\sst I}'{}\;\equiv\;\int_{I'}d\si\;\SE^b(\si).
\end{equation}
By using the formulae
\[
\begin{aligned}
\SQ_{\rm\sst I}'{}\SE_<^{\al_\1}(\si_\1)\;=& \;e^{-3\pi i b\al_\1}
\SE_<^{\al_\1}(\si_\1)\SQ_{\rm\sst I}'{},\\
\SQ_{\rm\sst I}^c{}\SE_<^{\al_\1}(\si_\1)\;=& \;e^{-\pi i b\al_\1}
\SE_<^{\al_\1}(\si_\1)\SQ_{\rm\sst I}^c{},
\end{aligned}
\qquad
\begin{aligned}
\SE_>^{\al_\2}(\si_\2)\SQ_{\rm\sst I}^c{}\;=& \;e^{\pi i b\al_\2}\SQ_{\rm\sst I}^c{}
\SE_>^{\al_\2}(\si_\2),\\
\SE_>^{\al_\2}(\si_\2)\SQ_{\rm\sst I}\;=& \;e^{-\pi i b\al_\2}\SQ_{\rm\sst I}
\SE_>^{\al_\2}(\si_\2),
\end{aligned}
\]
which follow easily from equation \rf{nof} one may represent
$\sh_{s_\2}^{\al_\2}(\si_\2)\sh_{s_\1}^{\al_\1}(\si_\1)$ by
\begin{equation}\begin{aligned}
{}& e^{\pi i \al_\2\al_\1}
|1-e^{i(\si_\1-\si_\2)}|^{2\al_\2\al_\1}
 \sh_{s_\2}^{\al_\2}(\si_\2)\sh_{s_\1}^{\al_\1}(\si_\1)\;=\;\\
& =\; \SE_<^{\al_\2}(\si_\2)\SE_<^{\al_\1}(\si_\1)
\Big[\big(\SQ_{\rm\sst I}^c{}+\SQ_{\rm\sst I}'{}e^{-2\pi ib\al_\1}\big)^{s_\2}
\big(\SQ_{\rm\sst I}^c{}+\SQ_{\rm\sst I}e^{-2\pi ib\al_\2}\big)^{s_\1}\Big]
\SE_>^{\al_\2}(\si_\2)\SE_>^{\al_\1}(\si_\1). 
\end{aligned}
\end{equation}
Let us next note that $\SQ_{\rm\sst I}^{}$, $\SQ_{\rm\sst I}^c{}{}$ and 
$\SQ_{\rm\sst I}'{}^{}$ satisfy algebraic relations of Weyl-type:
\begin{equation}\begin{aligned}
\SQ_{\rm\sst I}^c\SQ_{\rm\sst I}^{}\;=& \;q^{-2}\SQ_{\rm\sst I}^{}\SQ_{\rm\sst I}^c,\\
\SQ_{\rm\sst I}^c\SQ_{\rm\sst I}'\;=& \;q^{+2}\SQ_{\rm\sst I}'\SQ_{\rm\sst I}^c,
\end{aligned}\qquad \SQ_{\rm\sst I}^{}\SQ_{\rm\sst I}'\;=\;q^4\SQ_{\rm\sst I}'\SQ_{\rm\sst I}^{},
\end{equation}
where $q=e^{\pi i b^2}$. In order to continue we will
have to learn how to deal with complex powers of sums of 
Weyl-type operators.

\subsection{Interlude: Weyl-type operators}

What we have to deal with are expressions like
$(\SU+\SV)^s$, where $\SU$ and $\SV$ are positive self-adjoint operators
that satisfy $\SU\SV\;=\;q^{2}\SV\SU$.
Thanks to positivity we can take logarithms of $\SU$ and $\SV$ to 
express them as 
\[
\SU=e^{2b\sx},\qquad\SV=e^{b\sx}e^{2\pi b\spp}e^{b\sx},
\quad\text{where}\quad {[}\sx,\spp{]}= \frac{i}{2}.
\]
We would like to ``normal-order'' the expression
$(\SU+\SV)^s$, i.e. to write it in the
form $e^{bs\sx}f_s(\spp)e^{bs\sx}$.
The key ingredient for doing this will be 
the special function $e_b(z)$ which can be defined in terms of 
the function $s_b(x)$ (cf. the Appendix) by
$e_b(x)=e^{\frac{\pi i}{2}x^2}\,e^{-\frac{\pi i}{24}(2-Q^2)}s_b(x)$.
The following properties of $e_b(x)$ will be crucial for our purposes.
\begin{align}
{}& \text{(i) Functional equation:} \quad
e_b(x-i\fr{b}{2})=(1+e^{2\pi bx})e_b(x+i\fr{b}{2}). \label{e_feq}\\
{}& \text{(ii) Unitarity:}\qquad
|e_b(x)|=1\quad\text{for $x\in\BR$}. \label{e_unit}\\
{}& \text{(iii) Analyticity:}\quad
s_b(x)\;\text{is meromorphic,}\nonumber\\ 
{}& \hspace{2.7cm}\text{poles:}\;\,  
x=c_b+i(nb+mb^{-1}), n,m\in\BZ^{\geq 0}.\\
{}& \hspace{2.7cm}\text{zeros:}\;\,  
x=-c_b-i(nb+mb^{-1}), n,m\in\BZ^{\geq 0}.\nonumber 
\end{align}
With the help of these properties we may now calculate
\begin{equation}
\begin{aligned}
(\SU+\SV)^s\;=\;& \big(e^{b\sx}(1+e^{2\pi b\spp})e^{b\sx})^s\\
\={e_feq}\;& \bigg(e^{b\sx}
\frac{e_b(\spp-i\frac{b}{2})}{e_b(\spp+i\frac{b}{2})} 
e^{b\sx}\bigg)^s\;
=\; \big( e_b(\spp)e^{2b\sx}e_b^{-1}(\spp)\big)^s\\
\={e_unit}\;& e_b(\spp)e^{2sb\sx}e_b^{-1}(\spp)
\;=\; e^{sb\sx}\frac{e_b(\spp-is\frac{b}{2})}{e_b(\spp+is\frac{b}{2})}
e^{sb\sx}. 
\end{aligned}\label{WeylNO}
\end{equation}
On the right hand side we read off the desired normal ordering formula.

\subsection{Operator ordering, II}

Returning to the problem of the investigation of the conformal 
blocks let us now define operators $\sx$ and $\mss$
such that
\begin{equation}
\SQ_{\rm\sst I}^c{}\;=\;e^{2b\sx}, \qquad
\begin{aligned}
\SQ_{\rm\sst I}^{}\;=\;& e^{b\sx}e^{2\pi b (\mss-\spp)}e^{b\sx},\\
\SQ_{\rm\sst I}'{}\;=\;& e^{b\sx}e^{2\pi b (\mss+\spp)}e^{b\sx}.
\end{aligned}
\qquad [\sx,\spp]=\frac{i}{2},
\qquad 
\begin{aligned}
{[}\sx,\mss{]}=& 0,\\
[\spp,\mss]= & 0.
\end{aligned}
\end{equation}
It should be remarked that the operators $\sx$ and $\mss$ actually
depend on $\si_\1$ and $\si_\2$. We have suppressed this dependence
for notational convenience.
Generalizing the calculation in \rf{WeylNO} slightly
we find that 
\begin{equation}\begin{aligned}
\big(\SQ_{\rm\sst I}^c{}+e^{-2\pi i b\al_\1}\SQ_{\rm\sst I}'{}\big)^{s_\2}\;=\;&
e^{s_\2b\sx}\frac{e_b(\mss-i\al_\1+\spp-is_\2\frac{b}{2})}
{e_b(\mss-i\al_\1+\spp+is_\2\frac{b}{2})}e^{s_\2b\sx},\\
\big(\SQ_{\rm\sst I}^c{}+e^{-2\pi i b\al_\2}\SQ_{\rm\sst I}\big)^{s_\1}\;=\;&
e^{s_\1b\sx}\frac{e_b(\mss-i\al_\2-\spp-is_\1\frac{b}{2})}
{e_b(\mss-i\al_\2-\spp+is_\1\frac{b}{2})}e^{s_\1b\sx}.
\end{aligned}\end{equation}
Introducing the notation
\begin{equation}\begin{aligned}
\SE^{\al_\2\al_\1}_{s_\2s_\1}(\si_\2,\si_\1)&
\;\equiv\;
e^{b(s_\2+s_\1)\sx} \,\SE_>^{\al_\2}(\si_\2)\SE_>^{\al_\1}(\si_\1),\\
\big(\SE^{\al_\2\al_\1}_{s_\2s_\1}(\si_\2,\si_\1) & \big)^{\dagger}
\;\equiv\;
\SE_<^{\al_\1}(\si_\1)\SE_<^{\al_\2}(\si_\2)\,e^{b(s_\2+s_\1)\sx}
\end{aligned}\end{equation}
now leads to a representation for 
$\sh_{s_\2}^{\al_\2}(\si_\2)\sh_{s_\1}^{\al_\1}(\si_\1)$ of the
following form:
\begin{equation}\begin{aligned}
e^{\pi i \al_\2\al_\1}
|1-e^{i(\si_\1-\si_\2)}|^{2\al_\2\al_\1} & 
\sh_{s_\2}^{\al_\2}(\si_\2)\sh_{s_\1}^{\al_\1}(\si_\1)
\;=\;\\
=\;& \big(\SE^{\al_\2\al_\1}_{s_\2s_\1}(\si_\2,\si_\1)\big)^{\dagger}\;
\SO^{\al_\2\al_\1}_{s_\2s_\1}(\si_\2,\si_\1)\;
\SE^{\al_\2\al_\1}_{s_\2s_\1}(\si_\2,\si_\1)\;, 
\end{aligned}\label{cb_ordered}
\end{equation} 
where the operator $\SO^{\al_\2\al_\1}_{s_\2s_\1}(\si_\2,\si_\1)$ is defined
by 
\begin{equation}
\SO^{\al_\2\al_\1}_{s_\2s_\1}(\si_\2,\si_\1)\;\equiv
\frac{e_b(\mss+\spp-i\al_\1-i\frac{b}{2}(s_\2+s_\1))}
{e_b(\mss+\spp-i\al_\1+i\frac{b}{2}(s_\2-s_\1))}
\frac{e_b(\mss-\spp-i\al_\2-i\frac{b}{2}(s_\1+s_\2))}
{e_b(\mss-\spp-i\al_\2+i\frac{b}{2}(s_\1-s_\2))}
\end{equation}
We ultimately want to establish
a relation between the products
$\sh_{s_\2}^{\al_\2}(\si_\2)\sh_{s_\1}^{\al_\1}(\si_\1)$
and $\sh_{t_\1}^{\al_\1}(\si_\1)\sh_{t_\2}^{\al_\2}(\si_\2)$. 
Applying the same ordering procedure to the latter yields
\begin{equation}\begin{aligned}
e^{\pi i \al_\2\al_\1}
|1-e^{i(\si_\1-\si_\2)}|^{2\al_\2\al_\1} & 
\sh_{t_\1}^{\al_\1}(\si_\1)\sh_{t_\2}^{\al_\2}(\si_\2)
\;=\;\\
=\;& \big(\SE^{\al_\2\al_\1}_{t_\2t_\1}(\si_\2,\si_\1)\big)^{\dagger}\;
\SP^{\al_\2\al_\1}_{t_\2t_\1}(\si_\2,\si_\1)\;
\SE^{\al_\2\al_\1}_{t_\2t_\1}(\si_\2,\si_\1)
\end{aligned}
\end{equation} 
where now 
\begin{equation}
\SP^{\al_\2\al_\1}_{t_\2t_\1}(\si_\2,\si_\1)\;\equiv
\frac{e_b(\mss-\spp+i\al_\2-i\frac{b}{2}(t_\1-t_\2))}
{e_b(\mss-\spp+i\al_\2+i\frac{b}{2}(t_\2+t_\1))}
\frac{e_b(\mss+\spp+i\al_\1-i\frac{b}{2}(t_\2-t_\1))}
{e_b(\mss+\spp+i\al_\1+i\frac{b}{2}(t_\2+t_\1))}\;\, .
\end{equation}
We have put a lot of things into black boxes. The
definition of the operators $\mss$, for example, 
requires taking logarithms of the screening charges.
It may therefore 
seem far from clear that we have achieved anything useful
so far.  However, let us note that the operators
$\SE^{\al_\2\al_\1}_{s_\2s_\1}(\si_\2,\si_\1)$
are in fact {\it symmetric} w.r.t. exchange of the labels 1 and 2. 
The sought-for exchange relation \rf{braid} is therefore 
equivalent to an identity relating the operators 
$\SO^{\al_\2\al_\1}_{s_\2s_\1}(\si_\2,\si_\1)$
and $\SP^{\al_\2\al_\1}_{t_\2t_\1}(\si_\2,\si_\1)$.
Let us furthermore note that the 
two operators $\mss$ and $\spp$ commute with each other
and may therefore be simultaneously diagonalized. 
This implies that \rf{braid}
will be equivalent to an identity between 
ordinary meromorphic functions! 

\subsection{Projecting onto $\CF_{p_\2}$}


Let us now  consider the projection $\Pi_{p_\2}^{}
\sh_{s_\2}^{\al_\2}(\si_\2)\sh_{s_\1}^{\al_\1}(\si_\1)$. 
The operator $\spp$ can then be replaced by its eigenvalue 
$p_\2+\frac{i}{2}(\al_\2+\al_\1+bs_\2+bs_\1)$. For future
convenience let us introduce
$\mss\equiv\sr+\frac{i}{2}(\al_\2-\al_\1)$.
It will furthermore be convenient to trade the
parameters $(s_\2,s_\1,t_\2,t_\1,p)$ with $t_\1+t_\2=s_\2+s_\1$
for another set of parameters $(p_\2,p_\1,p_s,p_u)$ that is defined by
\begin{equation}\begin{aligned}
{} & p_\2=p, \qquad\;\;\; p_\1=p_\2+i(\al_\1+\al_\2+bs_\1+bs_\2),\\ 
{} & p_s=p_\1-i(\al_\1+bs_\1), \qquad p_u=p_\1-i(\al_\2+bt_\2).
\end{aligned}
\end{equation}
As a short notation for the
external parameters let us finally introduce
${\rm E}=(\al_\2,\al_\1,p_\2,p_\1)$.
 
$\Pi_{p_\2}^{}
\sh_{s_\2}^{\al_\2}(\si_\2)\sh_{s_\1}^{\al_\1}(\si_\1)$ and
$\Pi_{p_\2}^{}
\sh_{t_\1}^{\al_\1}(\si_\1)\sh_{t_\2}^{\al_\2}(\si_\2)$
can then be represented in the following form
\begin{equation}\label{projrepr}\begin{aligned}
e^{\pi i \al_\2\al_\1}
|1-e^{i(\si_\1-\si_\2)}|^{2\al_\2\al_\1} & \Pi_{p_\2}^{}
\sh_{s_\2}^{\al_\2}(\si_\2)  \sh_{s_\1}^{\al_\1}(\si_\1)\;=\;\\
=\;& \big(\SE^{\al_\2\al_\1}_{s_\2s_\1}(\si_\2,\si_\1)\big)^{\dagger}\;
\; O_{\rm\sst E}(\,p_s\,|\,\sr\,)\;
\SE^{\al_\2\al_\1}_{s_\2s_\1}(\si_\2,\si_\1)\;,\\
e^{\pi i \al_\2\al_\1}
|1-e^{i(\si_\1-\si_\2)}|^{2\al_\2\al_\1} & \Pi_{p_\2}^{}
\sh_{t_\1}^{\al_\1}(\si_\1)  \sh_{t_\2}^{\al_\2}(\si_\2)\;=\;\\
=\;& \big(\SE^{\al_\2\al_\1}_{t_\2t_\1}(\si_\2,\si_\1)\big)^{\dagger}\;
P_{\rm\sst E}(\,p_u\,|\,\sr\,)\;
\SE^{\al_\2\al_\1}_{t_\2t_\1}(\si_\2,\si_\1)\;,
\end{aligned}\end{equation}
where the operators $O_{\rm\sst E}(\,p_s\,|\,\sr\,)$ and
$P_{\rm\sst E}(\,p_u\,|\,\sr\,)$ are obtained from
$\SO^{\al_\2\al_\1}_{s_\2s_\1}(\si_\2,\si_\1)$
and $\SP^{\al_\2\al_\1}_{t_\2t_\1}(\si_\2,\si_\1)$ respectively
by making the substitutions described above.

Let us finally replace the special functions $e_b(x)$ 
by their close relatives $s_b(x)$
\begin{equation}
s_b(x)\;=\;
e^{-\frac{\pi i}{2}x^2}e^{\frac{\pi i}{24}(2-Q^2)}e_b(x),
\end{equation}
which satisfy an inversion relation of the simple form
\begin{equation}\label{invers}
s_b(x)s_b(-x)\;=\;1.
\end{equation}
We then find that $O_{\rm\sst E}(\,p_s\,|\,r\,)$ and $P_{\rm\sst E}(\,p_u\,|\,r\,)$
are given by expressions of the following form
\begin{equation}\begin{aligned}
O_{\rm\sst E}(\,p_s\,|\,r\,)\;=\;& C_{\rm\sst E}^s(p_s)
\,e^{\pi b(s_\1+s_\2)r}\;
F_{\rm\sst E}^s(\,p_s\,|\,r\,),\\
P_{\rm\sst E}(\,p_u\,|\,r\,)\;=\;& C_{\rm\sst E}^u(p_u)
\,e^{\pi b(s_\1+s_\2)r}\;
F_{\rm\sst E}^u(\,p_u\,|\,r\,),
\end{aligned}\end{equation}
where $F_{\rm\sst E}^{\flat}(\,p_s\,|\,r\,)$, $\flat\in\{s,u\}$ 
are defined respectively by
\begin{equation}\label{basis}\begin{aligned}
F_{\rm\sst E}^s(\,p_s\,|\,r\,)\;=\;& \frac{s_b(r+p_\2+i(\al_\2-\al_\1))s_b(r-p_\1)}{
s_b(r+p_s-i\al_\1)s_b(r-p_s-i\al_\1)} \\
F_{\rm\sst E}^u(\,p_u\,|-r\,)\;=\;& \frac{s_b(r+p_\2+i(\al_\1-\al_\2))s_b(r-p_\1)}{
s_b(r+p_u-i\al_\2)s_b(r-p_u-i\al_\2)}.
\end{aligned}\end{equation}
We have used the inversion relation \rf{invers}. 
Otherwise we will only need to know the ratio
$C_{\rm\sst E}^s(p_s)/C_{\rm\sst E}^{u}(p_u)$ which is found to be 
of the form $C_{\rm\sst E}(s)/D_{\rm\sst E}(t)=e^{\pi i(\De_s+\De_u-\De_\1-\De_\2)}$
where $\De_{\natural}\equiv p^2_{\natural}+\frac{1}{4}Q^2$, 
$\natural\in\{1,2,s,u\}$.
 
\subsection{The main identity}

Let us momentarily assume that $\al_k\in\frac{Q}{2}+i\BR$, $k=1,2$.
The functions
$F_{\rm\sst E}^{\flat}(\,p_s\,|\,r\,)$ have singularities 
as function of $r\in\BR$. However, the 
$F_{\rm\sst E}^{\flat}(\,p_s\,|\,r+i0\,)$ 
represent well-defined
distributions that will be denoted as $\bra\,{\rm d}_p^\flat\,|$ for 
$\flat=s,u$.
The crucial observation that finally leads us to the desired braid relations
\rf{braid} is now the following:

\begin{thm}\cite{PT2}
The following two sets of distributions
\[ 
{\mathfrak F}_{\rm\sst E}^{\flat}\;\equiv\;
\big\{\,|\,{\rm d}_p^\flat\,\ket \;;
\;p_s\in\BR^+\,\big\}, \qquad
\flat\in\{s,u\}
\]
form bases for 
$L^2(\BR)$
in the sense of generalized functions. We have the relations
\begin{equation} \label{orthcomp} \begin{aligned}
{} & \bra\,{\rm d}_p^\flat\,|\,{\rm d}_q^\flat\,\ket = \;{\rm m}^{-1}(p)\,\de(p-q)
\qquad\qquad\;\text{\rm (Orthogonality)}, \\
{}& 
\int_{\BR^+}dp \;{\rm m}(p) \;|\,{\rm d}_p^\flat\,\ket \bra\,{\rm d}_p^\flat\,|
\;= \;\id \qquad \qquad\text{\rm (Completeness)}.
\end{aligned}\end{equation} 
\end{thm} 
The proof of this result can be simplified considerably by noting
that the functions in the numerators of \rf{basis} are pure
phases due to the property $|s_b(x)|=1$ for $x\in\BR$ of the
function $s_b$. 
Multiplying a function $\psi(r)\in L^2(\BR)$
by these numerators therefore defines a unitary operator on $L^2(\BR)$.
Moreover, the numerators in \rf{basis} do not depend on the
labels $p_s$, $p_u$ of the elements of ${\mathfrak F}_{\rm\sst E}$. 
Let us furthermore note that translation 
invariance of $L^2(\BR)$ allows one to ignore 
the imaginary part of $\al_i$ in the denominators.
The statement of the theorem is therefore 
equivalent to the corresponding statement for the set of distributions
${\mathfrak F}\,\equiv\,
\big\{\,f(\,p_s\,|\,r+i0\,) \;;\;p_s\in\BR^+\big\}$, where
\[
f(\,p_s\,|\,r\,)\,=\, s_b^{-1}(r+p_s-c_b)s_b^{-1}(r-p_s-c_b)\;,
\]
where $c_b=iQ/2$.
There is a short and direct
proof for the completeness
of the set ${\mathfrak F}$ due to R. Kashaev \cite{Ka}.

We are now in the position to complete the proof of \rf{braid}.
The completeness relation in \rf{orthcomp} immediately implies that
\begin{equation}
\int_{\BR^+}dq \;\,|\,{\rm d}_{q}^u\,\ket\;B_{\rm\sst E}^{+}(\,p\,|\,q\,)\;
=\; |\,{\rm d}_{p}^s\,\ket,\qquad
B_{\rm\sst E}^{+}(\,p\,|\,q\,)\;\equiv\; {\rm m}(q)\,
\bra\,{\rm d}_q^u\,|\,{\rm d}_p^s\,\ket\;\, .
\end{equation}
This is equivalent to the identity
\begin{equation}
\int_{\BR^+}dp_u \;B_{\rm\sst E}^{+}(\,p_s\,|\,p_u\,)\,P_{\rm\sst E}
(\,p_u\,|\,r\,)\;
=\; O_{\rm\sst E}(\,p_s\,|\,\sr\,),
\end{equation}
which implies the desired braid-relation \rf{braid} thanks to
\rf{projrepr}. The generalization of this relation to more general
values of $\al_k$, $s_k$, $k=1,2$ can then be obtained by analytic
continuation.
The unitarity relations \rf{unit} also follow
easily from the completeness relations in \rf{orthcomp}.

\section{Fusion}

In order to have a more uniform
notation, we will in the following replace the labels $\al$ 
of the chiral vertex operators $\sh_{p_\2p_\1}^{\al}(z)$ by 
$q=c_b-i\al$, $c_b=i\frac{Q}{2}$.

\subsection{Descendants}

It is often useful to generalize the chiral vertex operators
$\sh_{p_\2p_\1}^{q}(z)$ by introducing a family
of operators $\sh_{p_\2p_\1}^{q}(\xi\,|\,z)$ which are labeled
by elements $\xi\in\CF_{q}$.
The operators $\sh_{p_\2p_\1}^{q}(\xi\,|\,z)$ are 
defined in terms of $\sh_{p_\2p_\1}^{q}(z)$ by 
means of the requirements 
\begin{equation}\label{descdef}
\begin{aligned}
\text{(i)}\quad & \sh_{p_\2p_\1}^{q}(L_{-n}\xi\,|\,z)=\big((n-2)!\big)^{-1}
:\big(\pa^{n-2}T(z)\big)\sh_{p_\2p_\1}^{q}(\xi\,|\,z):\;\,\text{for}\;n>1,\\
\text{(ii)}\quad & \sh_{p_\2p_\1}^{q}(L_{-1}\xi\,|\,z)\;=\;\pa_z^{}
\sh_{p_\2p_\1}^{q}(\xi\,|\,z),\\
\text{(iii)}\quad & \sh_{p_\2p_\1}^{q}(v_p\,|\,z)\;=\;
\sh_{p_\2p_\1}^{q}(z),
\end{aligned}
\end{equation}
where the normal ordering in (i) is defined the same way as in \rf{normord}.
The definition of the descendants goes back to \cite{BPZ}, the present
formulation is the one from \cite{TL}. 
For our purposes the main point to notice is that
the definition of descendants introduces a second way to compose
the chiral vertex operators which may be represented by the
notation 
\begin{equation}\label{2ndcomp}
\sh_{p_\2p_\1}^{p_t}\big(\,\sh_{p_tq_\1}^{q_\2}(z_\2-z_\1)v_{q_\1}^{}\,|\,z_\1\big )
\;.
\end{equation}
This expression is defined by expanding 
$\sh_{p_tq_\1}^{q_\2}(z_\2-z_\1)v_{q_\1}^{}$ as 
\begin{equation}\label{shv_exp}
\sh_{p_tq_\1}^{q_\2}(z_\2-z_\1)v_{q_\1}^{}\;=\;
(z_\2-z_\1)^{\De_{p_t}-\De_{q_\2}-\De_{q_\1}}\sum_{n=0}^{\infty}(z_\2-z_\1)^n\;
\xi_{p_t}^{(n)}(q_\1,q_\2),
\end{equation}
where $\xi_{p_t}^{(n)}(q_\1,q_\2)\in\CF_{p_t}$. Terms like
$\sh_{p_\2p_\1}^{p_t}\big(\,\xi_{p_t}^{(n)}(q_\1,q_\2)\,|\,z_\1\big )$ 
are defined by \rf{descdef}. We will see later that
the power series that is obtained by inserting \rf{shv_exp} 
into \rf{2ndcomp} has a finite radius of convergence within matrix
elements.

\subsection{Fusion - the result}

We shall now study the conformal blocks
\begin{equation}\begin{aligned}
\CF_{{\rm E},p_s}(z_\2,z_\1)\;\equiv\;&
\bra\, v_{p_\2}^{}\, ,\,
\sh_{p_\2p_s}^{q_\2}(z_\2)
\sh_{p_sp_\1}^{q_\1}(z_\1)\,v_{p_\1}^{}\,\ket\\
=:& \;z_\2^{\De_{p_\2}-\De_{p_\1}-\De_{q_\2}-\De_{q_\1}}
\CF_{{\rm E},p_s}(z), \qquad z\equiv\frac{z_\1}{z_\2}.
\end{aligned}\end{equation}
Our aim will be to 
describe the behavior of $\CF_{{\rm E},p_s}(z)$ near 
$z=1$. More precisely, our aim is to show that
$\CF_{{\rm E},p_s}(z)$ can be expanded as 
\begin{equation}\label{fusionrel}
\CF_{{\rm E},p_s}^s(z)\;=\;
\int\limits_{0}^{\infty}dp_t \;\,\Phi_{\rm\sst E}^{}(\,p_s\,|\,p_t)\;
\CF_{{\rm E},p_t}^t(z),
\end{equation}
where the ``t-channel'' conformal blocks $\CF_{{\rm E},p_t}^t(z)$
are defined by 
\begin{equation}
\CF_{{\rm E},p_t}^t(z)\;=\;\big\bra\,v_{p_\2}^{}\; ,\;
\sh_{p_\2p_\1}^{p_t}\big(\,\sh_{p_tq_\1}^{q_\2}(1-z)v_{q_\1}^{}\,|\,z\,\big )\,
v_{p_\1}^{}\,\big\ket\;,
\end{equation}
which implies existence of an expansion of the following form
\begin{equation}
\CF_{{\rm E},p_t}^t(z)\;=\;(1-z)^{\De_{p_t}-\De_{q_\2}-\De_{q_\1}}
\sum_{n=0}^{\infty}(1-z)^{n}\;\CF_{{\rm E},p_t}^t(n)\;\,.
\end{equation}
The ``fusion coefficients'' $\Phi_{\rm\sst E}^{}(\,p_s\,|\,p_t)$ are 
given as
 \begin{equation}\label{fuscoeff1}\begin{aligned}
\Phi_{\rm\sst E}^{}(\,p_s\,|\,p_t)\;=\;
\frac{s_b(w_\1)}{s_b(w_\2)}\frac{s_b(w_3)}{s_b(w_4)}\,
\frac{s_b(2q_\1-c_b)}{s_b(2p_u+c_b)}\,
\int\limits_{\BR}dt\;\,\prod_{i=1}^4\;\frac{s_b(t+u_i)}{s_b(t+v_i)},
\end{aligned}\end{equation}
where the coefficients $r_i$ and $s_i$, $i=1,\ldots,4$ are defined 
respectively by
\begin{equation}  
\begin{aligned} u_\1=& p_\1-q_\2-p_\2,\\
         u_\2=& p_\1-q_\2+p_\2, \\
        u_3=& +q_\1,\\
        u_4=& -q_\1,
\end{aligned}\qquad
\begin{aligned} 
       v_\1=& +p_u-q_\2-c_b,\\
         v_\2=& -p_u-q_\2-c_b,\\
         v_3=& +p_s+p_\1-c_b,\\
        v_4=& -p_s+p_\1-c_b. 
\end{aligned}\qquad
\begin{aligned} w_\1=& p_s+p_\1-q_\1,\\
         w_\2=& p_s+q_\1-p_\1, \\
        w_3=& p_\2+p_u-p_\1,\\
        w_4=& p_\2+p_\1-p_u.
\end{aligned}\end{equation}
The `fusion coefficients $\Phi_{\rm\sst E}^{}(\,p_s\,|\,p_t)$
are related to the b-Racah-Wigner coefficients $\big\{\dots\}_b^{}$
defined and studied in \cite{PT2} via
\begin{equation}\label{FRW}\begin{aligned}
\Phi_{\rm\sst E}^{}(\,p_s\,|\,p_t)\;=\;&
\frac{M(p_4,p_3,p_s)M(p_s,p_\2,p_\1)}{M(p_4,p_t,p_\1)M(p_t,p_3,p_\2)}
\;\Big\{\begin{smallmatrix} \al_\1 & \be_\1 \\[.5ex]
\bar{\be}_\2 & \bar{\al}_\2
\end{smallmatrix}\big|\begin{smallmatrix} \be_s\\[.5ex]
\bar{\be}_t \end{smallmatrix}\Big\}_b^{}.\\
=\;& \frac{M(p_4,p_3,p_s)M(p_s,p_\2,p_\1)}{M(p_4,p_t,p_\1)M(p_t,p_3,p_\2)}
\;\Big\{\begin{smallmatrix} \be_\1 & \al_\1 \\[.5ex]
\al_\2 & \be_\2
\end{smallmatrix}\big|\begin{smallmatrix} \be_s\\[.5ex]
\be_t \end{smallmatrix}\Big\}_b^{}.
\end{aligned}
\end{equation}
where we used the notation $\be_\flat=\frac{Q}{2}+ip_\flat$,
$\bar{\be}_\flat=\frac{Q}{2}-ip_\flat$ for $\flat=1,2,s,t$,
$\al_k=\frac{Q}{2}+iq_k$, $\bar{\al}_k=\frac{Q}{2}-iq_k$ for $k=1,2$ and 
\begin{equation}
M(p_3,p_\2,p_\1)\;=\;\frac{s_b(2p_\2-c_b)}{s_b(p_\2+p_3-p_\1)}\;.
\end{equation}
The first line in \rf{FRW} follows directly from the definitions, 
in order to go from the first to
the second line we have used a symmetry of the b-Racah-Wigner coefficients
found in \cite{PT3}, Appendix B.2.
This means that the b-Racah-Wigner coefficients describe the operator
product expansion of the rescaled fields $\sf_{p_3p_\1}^{p_\2}(w)=
(M(p_3,p_\2,p_\1))^{-1}\sh_{p_3p_\1}^{p_\2}(w)$.

\subsection{Elementary braid relation}

As a preparation 
we will need the limit of our general braid relation \rf{braid} where
$p_\1\ra c_b$. The representation $\CF_{c_b}$ is the vacuum representation
that has vanishing highest weight.
Let us observe that formula \rf{g-matel} implies that
both sides generically vanish in this limit unless some pole from the
$\Ga_b$-functions in the numerator cancels the zero from the factor
$1/\Ga_b(Q+2ip_\1)$. This is the case if one has set $p_s=p_\1$ 
before taking the limit, as we shall assume from now on. 
We are going to show that in the limit $p_\1\ra c_b$ the braid relation
\rf{braid} simplifies to 
\begin{equation}\label{elembraid}
\sh_{p_\2q_\1}^{q_\2}(\si_{\2})\,
\sh_{q_\1c_b}^{q_\1}(\si_{\1})\;=\;
\Omega^{\ep}(p_\2,q_\2,q_\1)\;
\sh_{p_\2q_\2}^{q_\1}(\si_{\1})\,\sh_{q_\2c_b}^{q_\2}(\si_{\2}),
\end{equation}
where $\ep=\sgn(\si_2-\si_1)$, 
\begin{equation}
\Omega^{\ep}(p_\2,q_\2,q_\1)\;=\;e^{\pi i\ep(\De_{q_\2}+\De_{q_\1}-\De_{p_\2})}
\frac{s_b(p_\2+q_\1-q_\2)}{s_b(p_\2+q_\2-q_\1)}
\frac{s_b(2q_\2-c_b)}{s_b(2q_\1-c_b)}.
\end{equation}

In order to derive \rf{elembraid} let us write $p_\1$ as
$p_\1=c_b-i\ep$ for a small positive $\ep$. It is easy to see that
the matrix
element $\langle  v_{p_u}\, ,\, \sh_{p_u p_\1}^{q_\2}(1)\,
v_{p_\1} \rangle$ behaves for $\ep\ra 0$
as \begin{equation}
\langle  v_{p_u}\, ,\, \sh_{p_u p_\1}^{q_\2}(1)\,
v_{p_\1} \rangle\;\underset{\ep\ra 0}{\sim}\;
\frac{2\ep}{\ep+i(p_u-q_\2)}\;.
\end{equation}
A non-vanishing integrand in \rf{braid} is therefore only found if 
$p_u=q_\2$. However, in order to extract the singular
behavior for $\ep\ra 0$, $p_u\ra q_\2$ we need to observe that the
braid coefficients \rf{braidcoeff1} develop a pole at these 
parameter values. This pole comes from the fact that the pole
at $t=c_b-r_3$ from the numerator and the pole at $t=-c_b-s_\1$ 
from the denominator of the integrand both approach the 
point $t=0$ in the limit $\ep\ra 0$, $p_u\ra q_\2$. In order to
extract the resulting singular behavior of the integral one may
deform the contour of integration into a contour that separates the
pole at $t=c_b-r_3$ from all the other poles in the upper t-half-plane,
and a small circle around the point $t=c_b-r_3$. Only the contribution
from the small circle shows singular behavior in the limit that
we consider. 
By taking into account eqn. \rf{sbRes}
we find that
\begin{equation}
\int_{\BR} dt\;\prod_{i=1}^4\frac{s_b(t+r_i)}{s_b(t+s_i)}\;
\underset{\begin{smallmatrix} p_u\ra q_\2 \\ \ep\ra 0
\end{smallmatrix}}{\sim}\;
\frac{s_b(p_\2+q_\1-q_\2)}{s_b(p_\2+q_\2-q_\1)}
\frac{s_b(2q_\2-c_b)}{s_b(2q_\1-c_b)}
\frac{1}{2\pi i}\frac{-1}{p_u-q_\2+i\ep}\;
\end{equation}
up to terms that are regular in this limit.
Collecting the $\ep$-dependent factors one finds
\[
\frac{1}{2\pi}
\frac{2\ep}{(p_u-q_\2)^2+\ep^2}\;\underset{\ep\ra 0}{\rightarrow}\;
\de(p_u-q_\2).
\]
To complete the verification of 
our claim \rf{elembraid} is now the matter of a 
straightforward calculation.

\subsection{Fusion - the derivation}

Let us consider the following matrix element:
\begin{equation}
\CF_{{\rm E},p_s}(z_\2,z_\1,z_0)\;\equiv\;\bra\, v_{p_\2}^{}\, ,\,
\sh_{p_\2p_s}^{q_\2}(z_\2)
\sh_{p_sp_\1}^{q_\1}(z_\1)
\sh_{p_\1c_b}^{p_\1}(z_0)\,v_0^{}\,\ket\;,
\end{equation}
where $v_0\equiv v_{c_b}$. We will assume 
for the moment that $z_k=e^{i\si_k}$, $k=0,1,2$ 
with $\si_\2>\si_\1>\si_0$.  
$\CF_{{\rm E},p_s}(z_\2,z_\1,z_0)$ is the boundary value of a function that
is analytic for $|z_\2|>|z_\1|>|z_0|$.
We may then use
the relations \rf{elembraid}
and \rf{braid} to express $\CF_{{\rm E},p_s}(z_\2,z_\1,z_0)$ as
\begin{equation}\label{doublebraid}\begin{aligned}
{}& \CF_{{\rm E},p_s}(z_\2,z_\1,z_0)\;\equiv\;\\[-.5ex]
& =\;\Omega^{+}(p_s,q_\1,p_\1)\;
\int\limits_{0}^{\infty}dp_u\;B_{\rm D}^+(\,p_s\,|\,p_u)
\;\bra\, v_{p_\2}^{}\, ,\,
\sh_{p_\2p_u}^{p_\1}(z_0)
\sh_{p_uq_\1}^{q_\2}(z_\2)
\sh_{q_\1c_b}^{q_\1}(z_\1)\,v_0^{}\,\ket\;,
\end{aligned}\end{equation}
where ${\rm D}=(q_\2,p_\1,p_\2,q_\1)$. Let us note the simple
relation $e^{z_\1L_{-1}}\xi=\sh_{p_us_b}^{p_u}(\,\xi\,|\,z_\1)v_0$,
which implies that $\sh_{p_uq_\1}^{q_\2}(z_\2)
\sh_{q_\1c_b}^{q_\1}(z_\1)\,v_0^{}$ can be written as 
\begin{equation}\label{VOid}
\begin{aligned}
\sh_{p_uq_\1}^{q_\2}(z_\2)
\sh_{q_\1c_b}^{q_\1}(z_\1)\,v_0^{}\;=\;&
e^{z_\1L_{-1}}\sh_{p_uq_\1}^{q_\2}(z_\2-z_\1)v_{q_\1}^{}\\
=\;& \sh_{p_us_b}^{p_u}\big(\,
\sh_{p_uq_\1}^{q_\2}(z_\2-z_\1)v_{q_\1}^{}\,|\,z_\1\big)\,v_0.
\end{aligned}
\end{equation}
By inserting \rf{VOid} into \rf{doublebraid} and using
\rf{elembraid} (now with $\ep=-1$) again we finally find the expression
\begin{equation}\begin{aligned}
{} \CF_{{\rm E},p_s} & (z_\2,z_\1,z_0)\;\equiv\;\\[-.5ex]
& =\;
\int\limits_{0}^{\infty}dp_u\;\Phi_{\rm E}(\,p_s\,|\,p_u)
\;\bra\, v_{p_\2}^{}\, ,\,
\sh_{p_\2p_\1}^{p_u}\bigl(\,\sh_{p_uq_\1}^{q_\2}(z_\2-z_\1)v_{q_\1}^{}\,|\,z_\1
\,\bigr)\sh_{p_uc_b}^{p_u}(z_0)
\,v_0^{}\,\ket\;,
\end{aligned}\end{equation}
where the coefficients $\Phi_{\rm E}(\,p_s\,|\,p_u)$ are given 
as
\begin{equation}
\Phi_{\rm E}(\,p_s\,|\,p_u)\;=\;
\Omega^{+}(p_s,q_\1,p_\1)\,B_{\rm D}^+(\,p_s\,|\,p_u)\,
\Omega^{-}(p_\2,p_\1,p_u)\;.
\end{equation}
By assembling the pieces one verifies that this expression coincides 
with the one given in \rf{fuscoeff1}. 
The fact that the expansion in powers of $\frac{z_\2-z_\1}{z_\1-z_0}$ 
has a finite radius 
of convergence can be seen by returning to \rf{doublebraid} and noting that
\[ 
\bra\, v_{p_\2}^{}\, ,\,
\sh_{p_\2p_u}^{p_\1}(z_0)
\sh_{p_uq_\1}^{q_\2}(z_\2)
\sh_{q_\1c_b}^{q_\1}(z_\1)\,v_0^{}\,\ket\;=\;
\bra\, v_{p_\2}^{}\, ,\,
\sh_{p_\2p_u}^{p_\1}(z_0-z_\1)
\sh_{p_uq_\1}^{q_\2}(z_\2-z_\1)
\,v_{q_\1}^{}\,\ket\;,
\]
cf. our discussion in Subsection 2.4. 
It can finally be checked that the result is independent of our
choice of the order of the variables $\si_k$, $k=0,1,2$ 
if one excludes that case that $\si_0$ lies between $\si_\1$ and $\si_\2$.
The latter case is not suitable for studying the behavior where
$\si_\2$ approaches $\si_\1$.

\section{General exponential operators}\label{genexps}

\subsection{Introducing the right-movers} \label{rmov}

Let us now introduce a second, right-moving chiral field 
$\bar{\vf}(x_-)=\bar{\sq}+\bar{\spp} x_- + 
\bar{\vf}^{}_<(x_{-})+\bar{\vf}^{}_>(x_{-})$.
The commutation relations for the modes are obtained from
\rf{ccr} by replacing $(\sq,\spp,\sa_n)\ra (\bar{\sq},\bar{\spp},\ba_n)$.
The corresponding Hilbert-space will be denoted 
by $\CH^{\rm\sst F}_{\rm\sst R}$. Normal ordered exponentials $\bar{\SE}^{\al}(x_-)$
and 
screening charges $\bar{\SQ}(x_-)$
are defined by the obvious substitutions. This will {\it not} be 
case for our definition of $\bar{\sh}_s^{\al}(x_-)$, however. Instead we will
find it convenient to define $\bar{\sh}_s^{\al}(x_-)$ as
\begin{equation}
\bar{\sh}_s^{\al}(x_-)\;=\;\bar{\SS}^{-1}\,
\bar{\SE}^{\bar{\al}}(x_-)\,
\big(\bar{\SQ}(x_-)\big)^{\bar{s}}\,\bar{\SS}
\;,\end{equation}
where $\bar{\al}=Q-\al$, $\bar{s}=-s-Q/b$, and $\bar{\SS}$ is the 
intertwining operator that is defined in the same way as 
we have defined $\SS$ in Subsection \ref{confsym}. The definition
is such that the fields $\bar{\sh}_s^{\al}(\bw)$ have 
the shift property 
\begin{equation}\label{shift'}
\bar{\sh}_s^{\al}(\bw)f(\bar{\spp})\;=\;f\big(\bar{\spp}-i(\al+bs)\big)
\bar{\sh}_s^{\al}(\bw).
\end{equation}

The formula for the matrix elements of the fields 
$\bar{\sh}_s^{\al}(x_-)$ is obtained
from \rf{g-matel} by the substitutions $s\ra -s-Q/b$, $p_i\ra -p_i$ for 
$i=1,2$ as well as $\al\ra Q-\al$. In order to find the exchange relations 
of the fields $\bar{\sh}_s^{\al}(x_-)$ from \rf{braid} it is enough to
replace $\al_i\ra Q-\al_i$ and $\si_i\ra -\si_i$ 
for $i=1,2$. 

In the following we will only be interested in the 
diagonal subspace $\CH^{\rm\sst F}_{\rm \sst D}$ 
in $\CH^{\rm\sst F}\;=\;
\CH^{\rm\sst F}_{\rm\sst L}\ot
\CH^{\rm\sst F}_{\rm\sst R}$
that is defined by the condition $\spp=\bar{\spp}$. 
We clearly have $\CH^{\rm\sst F}\;\simeq\;
L^2(\BR)\ot\CF\ot \CF$. 
We shall only consider 
combinations of the chiral
fields like
$\sh_s^{\al}(x_+)\ot\bar{\sh}_s^{\al}(x_-)$.
These fields preserve the diagonal $\spp=\bar{\spp}$
thanks to the equality of 
shifts \rf{shift} and \rf{shift'}. The projection to 
$\CH^{\rm\sst F}_{\rm \sst D}$ is automatic if one considers matrix elements
like $\bra v_{p}\ot \bar{v}_{p} , \SO\,\psi\ket_{\CH^{\rm\sst F}}^{}.$

\subsection{Construction}

Let us introduce the following one-parameter family of fields
\begin{equation}\label{Vfields}
\boxed{\quad
\begin{aligned}
\SV_{\al}(w,\bw)\;\equiv\;& \la_b\,n(\al)
\;\SN^{-1} \;\SW_{\al}(w,\bw) \;\SN,\\
\SW_{\al}(w,\bw)\;\equiv\;& \;\frac{1}{2}\int_{\BT}ds
\;\,{\rm m}(\spp)\;\sh_s^{\al}(w)\ot\bar{\sh}_s^{\al}(\bar{w}).
\end{aligned}\quad}
\end{equation}
Besides the notation $\BT\equiv -\frac{Q}{2}+i\BR$ 
we have introduced the following objects in \rf{Vfields}:
\begin{align} 
{}& \text{(i) Field normalization factors:}&  & 
\left\{\;\begin{aligned} {} & n(\al)\;\equiv\;
\big(\pi \mu \ga(b^2)b^{2-2b^2}\big)^{-\frac{\al}{b}}
\frac{\Ga_b(2\al-Q)}{\Ga_b(2\al)}, \\
& \la_b\;=\; 
2\pi \big(\Ga(b^2)b^{1-b^2}\big)^{\frac{1+b^2}{b^2}},
\end{aligned}\right.
\label{normfac1}\\
{}& \text{(ii) Similarity transformation:}& & \SN\equiv N(\spp)\equiv 
\big(\pi \mu \ga(b^2)b^{2-2b^2}\big)^{-\frac{i}{b}\spp}
\frac{\Ga_b(Q-2i\spp)}{\Ga_b(-2i\spp)}
\label{normfac2}
\end{align}
The field $\SV_{\al}(w,\bw)$ represents the exponentials of the
Liouville field in the quantized theory.

\subsection{Properties}

The following properties of the fields $\SV_{\al}(w,\bw)$
can be seen to capture the essence of the exact solution
of the quantized Liouville theory \cite{TL}.\\[1ex]
{\sc Covariance} $\frac{\quad}{}$
\begin{equation}
\begin{aligned}
{[}\,\SL_n\, ,\, \SV_{\al}(w,\bw){]}\;=\;& 
    e^{nw}(\pa_w+\De_{\al}n)\SV_{\al}(w,\bw),\\
{[}\,\bar{\SL}_n\, ,\, \SV_{\al}(w,\bw){]}\;=\;&
    e^{n\bw}(\pa_{\bw}+\De_{\al}n)\SV_{\al}(w,\bw).
\end{aligned}
\end{equation}
{\sc Matrix elements} $\frac{\quad}{}$
\newcommand{\Up}{\Upsilon}
\begin{equation}\label{V-matel}
\begin{aligned}
{} & e^{-(w+\bw)(\De_{p_\2}-\De_{p_\1})}\;
\langle\,  v_{p_\2}\!\ot \bar{v}_{p_\2}\, ,\,\SV_{\al}(w,\bw) \,
v_{p_\1}\!\ot \bar{v}_{p_\1}\, \rangle\;=\;\\
& \qquad\qquad =\;  \big(\pi\mu\ga(b^2)b^{2-2b^2}\big)^s\;
\frac{\Up_0
\Up_b(Q-2ip_\2)\Up_b(2\al)\Up_b(Q+2ip_\1)}{|\, \Up_b(\al+i(p_\2+p_\1))
\Up_b(\al+i(p_\2-p_\1))\,|^2}
\end{aligned}
\end{equation}
We have again assumed $\al\in\BR$ in order to write the 
expression compactly. The special function 
$\Up_b(x)$ is defined\footnote{This definition differs from the 
one used in \cite{ZZ} by a b-dependent factor that
drops out in \rf{V-matel}} as $\Up_b^{-1}(x)=\Ga_b(x)\Ga_b(Q-x)$,
and $\Up_0=\2\pi\Ga_b^{-2}(Q)$. \\[1ex]
{\sc Locality} $\frac{\quad}{}$
\begin{equation}\label{locality}
\Pi_{\rm D}^{}\;\big[\,\SV_{\al_\2}(\si_\2)\, , \,
\SV_{\al_\1}(\si_\1) \big]\;=\;0\quad\text{where}\;\, \SV_{\al}(\si)\equiv
 \SV_{\al}(w,\bw)|_{w=i\si}^{}\;\, ,  
\end{equation}
and $\Pi_{\rm D}^{}: \CH^{\rm\sst F}_{}\ra
\CH^{\rm\sst F}_{\rm \sst D}$ denotes the projection to the diagonal
$\spp=\bar{\spp}$.

The calculation of the matrix elements \rf{V-matel} is straightforward.
One just
needs to combine the definitions of the previous subsection with
formula \rf{g-matel}, keeping in mind that 
one has to make the substitutions $\al\ra Q-\al$, $p_i\ra -p_i$ 
to get the matrix elements of the right-movers. 

In order to verify locality it clearly suffices to consider the fields
$\SW_{\al}$. Locality of these fields follows easily by first using 
the exchange relations \rf{braid} to change the order of the 
chiral components, and then simplifying the resulting expression
with the help of \rf{unit}. This works straightforwardly as long as 
$\al_k\in\frac{Q}{2}+i\BR$, $k=1,2$, since the 
exchange relations of the fields $\bar{\sh}^{\al}_s$
then involve the complex conjugate of the kernel $B_{\rm\sst E}^{\ep}$. 
Validity of \rf{locality} for more general
values of $\al_k$ follows by analytic continuation.
Let us furthermore notice that conformal covariance is manifest
in our construction. 

\subsection{Final remarks}

It may be interesting to observe that our ansatz \rf{Vfields}
yields local covariant fields for {\it arbitrary} choice of the 
functions $n(\al)$ and $N(p)$. This freedom is eliminated 
by invoking further consistency conditions of the conformal bootstrap.
In order to see this, 
let us recall some elements of the discussion in \cite{TL}, Part I.
State-operator correspondence implies 
a relation between vacuum expectation values and matrix elements, 
\begin{equation}\begin{aligned}
\langle\,   v_{p_3}\!\ot  \bar{v}_{p_3}\,& ,\,\SV_{\al_\2}(z_\2,\bz_\2) \,
v_{p_\1}\!\ot \bar{v}_{p_\1}\, \rangle \;=\;\\
& =\;
\lim_{z_3\ra\infty}\lim_{z_\1\ra 0}\;|z_3|^{4\De_{\al_3}}
\langle 0\,|\,\SV_{\al_3}(z_3,\bz_\2)\SV_{\al_\2}(z_\2,\bz_\2)
\SV_{\al_\1}(z_\1,\bz_\1) \,|\,0
\rangle\\
& =:\;|z_\2|^{2(\De_{\al_3}-\De_{\al_\2}-\De_{\al_\1})}\,C(\al_3,\al_\2,\al_\1)
\end{aligned}\end{equation}
where $\al_\1=\frac{Q}{2}+ip_\1$ and $\al_3=\frac{Q}{2}-ip_3$. Locality
implies that the three point function
$C(\al_3,\al_\2,\al_\1)$ must be symmetric in its three arguments.
Locality of the fields $\SV_{\al}$ together with symmetry of the
three point functions then imply the crossing symmetry of the
four-point functions, see \cite{TL}, Section 5.

It is easy to show that the requirement that 
$C(\al_3,\al_\2,\al_\1)$ should be symmetric 
fixes 
the functions $n(\al)$ and $N(p)$ up to a constant, which may be re-absorbed 
into the definition of $\la_b$. 
The left-over freedom 
is on the one hand the possibility to redefine
the prefactor $\la_b$, and on the other 
hand the freedom to replace the constant $\mu$ by a function of $b$.
This remaining freedom can be fixed by demanding 
\begin{itemize}\item[(i)] 
that
$\lim_{\al\ra 0}\SV_{\al}=\id$, which implies \cite{TL}, Section 4
\begin{equation}
\lim_{\al\ra 0}C\big(\fr{Q}{2}-ip_\2,\al,\fr{Q}{2}+ip_\1\big)\;=\;
2\pi\de(p_\2-p_\1)\, ,
\end{equation}
\item[(ii)] that the quantized Liouville field 
$\phi(w,\bw)=\big(\frac{1}{2}\pa_{\al}\SV_{\al}(w,\bw)\big)^{}_{\al=0}$
satisfies a quantum version of the Liouville 
equation which takes 
the specific form \cite{TL} (Section 9)
\begin{equation}
\pa_w\bar{\pa}_{\bw}\phi(w,\bw)\;=\;\pi \mu b\;\SV_{b}(w,\bw).
\end{equation}
\end{itemize}
If we adopt these requirements we are indeed forced to choose $n(\al)$, 
$N(p)$ and $\la_b$ as given in \rf{normfac1}, \rf{normfac2} 
The resulting formula for the
three point function $C(\al_3,\al_\2,\al_\1)$ is exactly the 
one proposed in \cite{ZZ}. 

\bigskip
\noindent{\bf Acknowledgments} $\frac{\quad}{}$
The author would like to thank the organizers of the 6th 
International Conference on CFTs and
Integrable Models, Chernogolovka, for organizing a nice and
stimulating conference. 
I am furthermore grateful to the SPhT CEA-Saclay
where this work was completed for hospitality. 
I would finally like to acknowledge support by the SFB 288 of the DFG.

\appendix
\section{Special functions}

\subsection{The function $\Ga_b(x)$}

The function $\Ga_b(x)$ is a close relative of the double
Gamma function studied in \cite{Ba,Sh}. It 
can be defined by means of the integral representation
\begin{equation}
\log\Ga_b(x)\;=\;\int\limits_0^{\infty}\frac{dt}{t}
\biggl(\frac{e^{-xt}-e^{-Qt/2}}{(1-e^{-bt})(1-e^{-t/b})}-
\frac{(Q-2x)^2}{8e^t}-\frac{Q-2x}{t}\biggl)\;\;.
\end{equation}
Important properties of $\Ga_b(x)$ are
\begin{align}
{}& \text{(i) Functional equation:} \quad
\Ga_b(x+b)=\sqrt{2\pi}b^{bx-\frac{1}{2}}\Ga^{-1}(bx)\Ga(x). \label{Ga_feq}\\
{}& \text{(ii) Analyticity:}\quad
\Ga_b(x)\;\text{is meromorphic,}\nonumber\\ 
{}& \hspace{2.5cm}\text{poles:}\;\,  
x=-nb-mb^{-1}, n,m\in\BZ^{\geq 0}.\\
{}& \text{(iii) Self-duality:}\quad \Ga_b(x)=\Ga_{1/b}(x). \label{self-dual} 
\end{align}

\subsection{The function $s_b(x)$}

The function $s_b(x)$ may be defined in terms of 
$\Ga_b(x)$ as follows
\begin{equation}\label{sbdef}
s_b(x)\;=\;\Ga_b\big(\fr{Q}{2}+ix\big)\,/\,\Ga_b\big(\fr{Q}{2}-ix\big)\;.
\end{equation}
This function, or close relatives of it like 
$e_b(x)\;=\;e^{\frac{\pi i}{2}x^2}\,e^{-\frac{\pi i}{24}(2-Q^2)}s_b(x)$,
have appeared in the literature under various names like 
``Quantum Dilogarithm'' \cite{FK1}, ``Hyperbolic G-function''
\cite{Ru}, ``Quantum Exponential Function'' \cite{W} 
and ``Double Sine Function'', we refer to 
the appendix of \cite{KLS} for
a useful collection of properties of $s_b(x)$ and further references.
An integral that represents $\log s_b(x)$ is
\begin{equation}
\log s_b(x)\;=\;\frac{1}{i}\int\limits_0^{\infty}\frac{dt}{t}
\biggl(\frac{\sin 2xt}{2\sinh bt\sinh b^{-1}t}-
\frac{x}{t}\biggl)\;\;.
\end{equation}
The most important properties for our purposes are 
\begin{align}
{}& \text{(i) Functional equation:} \quad
s_b\big(x-i\fr{b}{2}\big)\;=\;2\cosh \pi b x\;
s_b\big(x+i\fr{b}{2}\big). \label{sb_feq}\\
{}& \text{(ii) Analyticity:}\quad
s_b(x)\;\text{is meromorphic,}\nonumber\\ 
{}& \hspace{2.7cm}\text{poles:}\;\,  
x=c_b+i(nb-mb^{-1}), n,m\in\BZ^{\geq 0}.\\
{}& \hspace{2.7cm}\text{zeros:}\;\,  
x=-c_b-i(nb-mb^{-1}), n,m\in\BZ^{\geq 0}.\nonumber \\
{}& \text{(iii) Self-duality:}\quad s_b(x)=s_{1/b}(x).  \\
{}& \text{(iv) Inversion relation:}\quad s_b(x)s_b(-x)\;=\;1.\\
{}& \text{(v) Unitarity:} \quad \overline{s_b(x)}\;=\;1/s_b(\bar{x}).\\
{}& \text{(vi) Residue:} \quad {\rm res}_{x=c_b}s_b(x)
=(2\pi i)^{-1}\label{sbRes}.
\end{align}

\end{document}